\documentclass[epjST]{svjour}

\usepackage{graphics}

\usepackage[colorlinks=true,citecolor=blue,linkcolor=red,linktocpage=true,pagebackref=false]{hyperref}
\usepackage{amsmath, amsfonts, amssymb}
\usepackage{color}
\usepackage{braket}

\begin{document}

\title{
Composite two-particle sources
}
\author{Michael Moskalets\inst{1}\fnmsep\thanks{\email{michael.moskalets@gmail.com}}
\and Janne Kotilahti\inst{2}
\and Pablo Burset\inst{2}
\and Christian Flindt\inst{2}
}
\institute{Department of Metal and Semiconductor Physics, NTU ``Kharkiv Polytechnic Institute", 61002 Kharkiv, Ukraine
\and
Department of Applied Physics, Aalto University, 00076 Aalto, Finland
}

\abstract{Multi-particle sources constitute an interesting new paradigm following the recent development of on-demand single-electron sources. Versatile devices can be designed using several single-electron sources, possibly of different types, coupled to the same quantum circuit.
However, if combined \textit{non-locally} to avoid cross-talk, the resulting architecture becomes very sensitive to electronic decoherence.
To circumvent this problem, we here analyse two-particle sources that operate with several single-electron (or hole) emitters attached in series to the same electronic waveguide. We demonstrate how such a device can emit exactly two electrons without exciting unwanted electron-hole pairs if the driving is adiabatic. Going beyond the adiabatic regime, perfect two-electron emission can be achieved by driving two quantum dot levels across the Fermi level of the external reservoir. If a single-electron source is combined with a source of holes, the emitted particles can annihilate each other in a process which is governed by the overlap of their wave functions. Importantly, the degree of annihilation can be controlled by tuning the emission times, and the overlap can be determined by measuring the shot noise after a beam splitter. In contrast to a Hong-Ou-Mandel experiment, the wave functions overlap close to the emitters and not after propagating to the beam splitter, making the shot noise reduction less susceptible to electronic decoherence.
}

\maketitle

\section{Introduction}
\label{intro}

On-demand single-electron sources~\cite{Bauerle:2018ct,Splettstoesser:2017jd,Pekola:2013fg,Giblin:2019vv} make it possible to perform quantum optics experiments with electrons, and they form the basis of the rapidly developing field of {\it quantum coherent electronics}~\cite{Splettstoesser:2009im,Bocquillon:2013fp}. Electronic counterparts of the famous Hanbury-Brown and Twiss effect~\cite{HanburyBrown:1956bi,Dubois:2013ul,Bocquillon:2012if,Glattli:2017vp,Henny:1999tb,Oliver:1999ws,Oberholzer:2000wx,Neder:2007jl} and Hong-Ou-Mandel interference~\cite{Dubois:2013ul,Hong:1987gm,Freulon:2015jo,Bocquillon:2013dp,Marguerite:2016ur,Glattli:2016wl,Marguerite:2016jt} have already been demonstrated experimentally. Quantum tomography of single-electron states has also been proposed and realized~\cite{Grenier:2011dv,Jullien:2014ii,Marguerite:2017tn,Fletcher:2019vf}. However, one challenging issue working with electrons is decoherence and relaxation caused by neighboring electrons in the waveguides, which can significantly degrade quantum effects, resulting for example in a non-perfect Pauli dip in the shot noise (the fermionic analogue of the Hong-Ou-Mandel peak in quantum optics)~\cite{Bocquillon:2013dp}.
Several methods are currently being developed to prevent decoherence, including quantum environment engineering~\cite{Altimiras:2010dk,Huynh:2012tl,Cabart:2018ef,Rodriguez:2019wo,Duprez:2019wy}, and
the emission of electrons high above the Fermi level~\cite{Fletcher:2019vf,Kataoka:2015ta,Johnson:2018in}. Still, recent experimental results on energy relaxation indicate that the longest mean free path is achieved by emitting the electrons close to the Fermi level~\cite{Rodriguez:2019wo}. For quantum information processing, it may thus be favorable to use single-electron sources that emit particles right on top of the Fermi sea, for instance, a quantum capacitor~\cite{Feve:2007jx,Gabelli:2006eg,Buttiker:1993wh,Keeling:2008ft}, a leviton source~\cite{Dubois:2013ul,Levitov:1996ie,Ivanov:1997wz,Keeling:2006hq}, or the recently proposed emitter based on time-dependent local gating~\cite{Misiorny:2017ua}.

\begin{figure}[t]
\centering
	\resizebox{0.99\columnwidth}{!}{\includegraphics{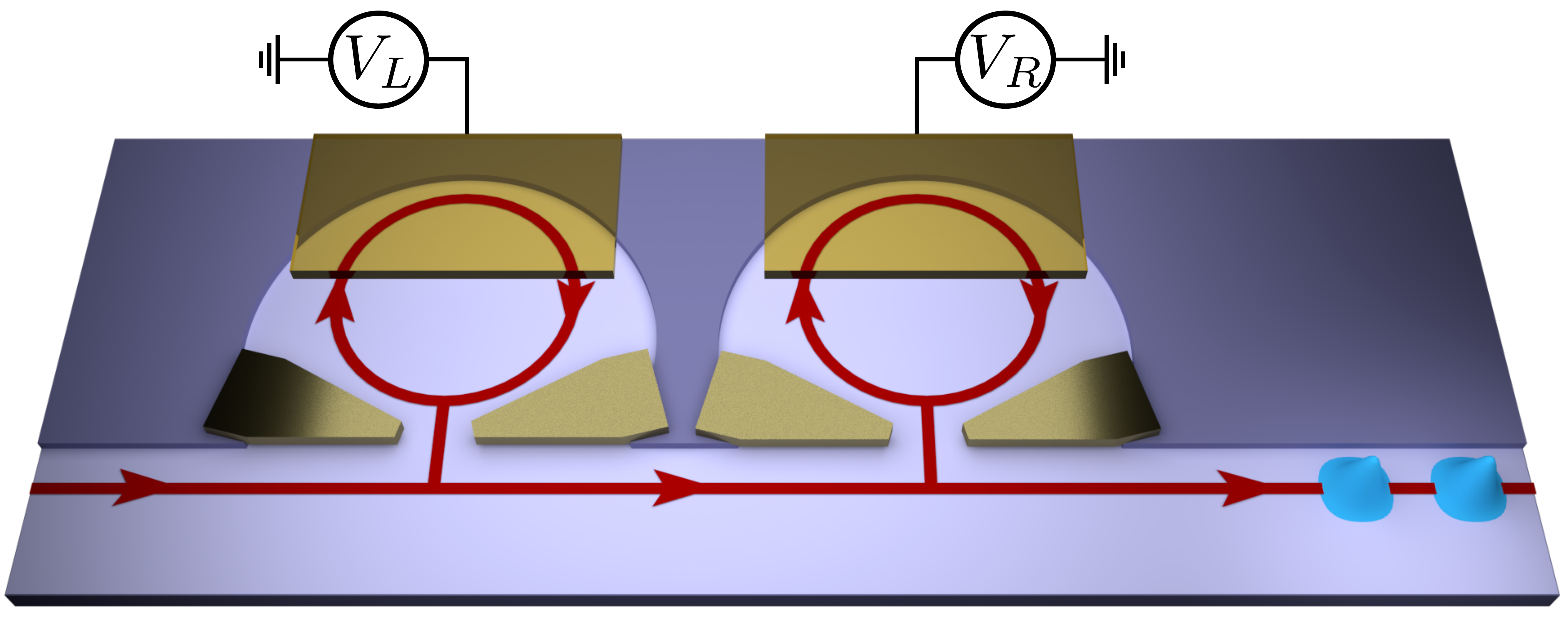} }
	\caption{A composite two-electron source can be constructed from single-electron sources located side by side. Here, two quantum capacitors in series are driven by separate gate voltages. Each capacitor is formed by a circular edge state (in red) and a metallic top gate. Electrons (in blue) are emitted into an edge state in the direction indicated by arrows.
	}
	\label{fig1}
\end{figure}

While single-electron emitters have been investigated both experimentally and theoretically, much less is known about more complex devices such as multi-particle sources, although some experimental progress in this direction has already been reported~\cite{Dubois:2013ul,Glattli:2016tr,Waldie:2015hy,Ubbelohde:2014vx,Fletcher:2012te,Fricke:2013cc}. In this work, we ee two-electron sources consisting of several single-electron sources connected in series as illustrated in Fig.~\ref{fig1}~\cite{Splettstoesser:2008gc}. Here we show two quantum capacitors attached to the same edge state, but one may also consider combinations of different types of emitters, for example by integrating a quantum capacitor~\cite{Feve:2007jx} with a source of levitons~\cite{Dubois:2013ul}. Importantly, as we will discuss in detail, one can combine a single-electron emitter with a source of holes~\cite{Freulon:2015jo}, which will enable the controlled creation of superpositions of quantum states which have different numbers of fermions. Note also that Coulomb interactions can presumably affect the emission process itself~\cite{Litinski:2016wp,Wagner:2019dq}, which has yet to be confirmed experimentally. Here, we do not take this effect into account.

The rest of the paper is organized as follows: In Sec.~\ref{s1}, we provide a brief account of the excess correlation function formalism and demonstrate its usefulness for the analysis of single and few-particle emitters. In Sec.~\ref{s2}, we outline the theory of composite two-electron sources based on the scattering properties of the individual single-particle emitters. We provide several examples of composite two-particle emitters in Sec.~\ref{s3}, and show that it is possible to emit exactly two electrons without exciting unwanted electron-hole pairs if the driving is adiabatic. In a specific case, perfect two-electron emission can also be achieved in the non-adiabatic regime. Technical details of our calculations are deferred to the Supplementary Material.

\section{Excess correlation function}
\label{s1}

The first-order correlation function is defined as ${\cal G} ^{(1)}\left( 1;2 \right) \!=\! \langle \hat\Psi^{\dag}_{ }(1)  \hat\Psi_{ }(2) \rangle_{}$, where $ \hat \Psi$ in our case is the field operator of electrons in the conductor of interest, and the brackets $\left \langle \dots \right \rangle$ denote the quantum-statistical average. The correlation function can be calculated using any many-body formalism, and in Sec.~\ref{s2} we show in particular how it can be evaluated using scattering theory.

The correlation function ${\cal G} ^{(1)}$ is additive in the number of electrons, which makes it a useful theoretical tool to characterize the excitations emitted by a source into a conductor. We just need to calculate ${\cal G} ^{(1)}$ with the source turned on and off and evaluate the difference known as {\it the excess first-order correlation function}~\cite{Grenier:2011dv,Grenier:2011js,Roussel:2017hu}
\begin{eqnarray}
G ^{(1)}_{}\left( 1;2 \right) = {\cal G} ^{(1)}_\text{on}\left( 1;2 \right) - {\cal G} ^{(1)}_\text{off}\left( 1;2 \right) .
\label{01}
\end{eqnarray}

To be specific, we consider a chiral one-dimensional conductor and fix the spatial coordinates at an arbitrary position downstream from the source, $x^{(0)}_{1} = x^{(0)}_{2} = x_{D}$, keeping only the times $t_{1}$ and $t_{2}$. For a linear (or linearized) dispersion relation, $ E - \mu = v_{ \mu} \left(  p - p_{ \mu} \right)$, the correlation function at different coordinates can be calculated using the substitution, $t_{ j} \to t_{ j} - v_{ \mu} \left(  x_{ j} - x_{D} \right)$, with $j=1,2$, where $\mu$ is the Fermi energy of the electrons in the conductor and $v_{\mu}$ ($p_{ \mu}$) is the velocity (momentum) of electrons at energy $E = \mu$. We restrict ourselves to  non-interacting electrons, where $G ^{(1)}_{}$ provides complete information about the injected particles, and we set the temperature to zero to keep the discussion simple.

\subsection{Single-particle emission}

If the source emits a single particle, the excess first-order correlation function reads~\cite{Grenier:2013gg}
\begin{eqnarray}
G ^{(1)}_{}\left( t_{1};t_{2} \right) = \eta
\frac{ e^{\frac{ i }{ \hbar } \mu \left(  t_{1} - t_{2} \right) } }{ v_{ \mu} }
\psi^{*}\left( t_{1}  \right) \psi^{}\left( t_{2}  \right),
\label{02}
\end{eqnarray}
where $ \eta = +1$ stands for the injection of an electron and $ \eta=-1$ for a hole.

The wave function is normalized such that
\begin{eqnarray}
\int _{}^{ } dt \left | \psi\left(  t \right) \right |^{2} = 1,
\label{03}
\end{eqnarray}
having introduced the factor of $1/ v_{ \mu}$ in Eq.~(\ref{02}), so that we can formulate the normalization condition as an integral over time, instead of over space. It turns out to be convenient to introduce the formal factor $e^{\frac{ i }{ \hbar } \mu \left(  t_{1} - t_{2} \right) }$ to describe the injection of an electron on top of the Fermi sea of the conductor, for example for a leviton source~\cite{Moskalets:2015kx}.

The correlation function is idempotent in the sense that
\begin{eqnarray}
\int _{}^{ } dt G ^{(1)}_{}\left(  t_{1}; t \right) G ^{(1)}_{}\left(  t; t_{2} \right)
=
\eta G ^{(1)}_{}\left(  t_{1}; t_{2} \right),
\label{04}
\end{eqnarray}
which is characteristic for a pure state. A discussion of single-particle emission at non-zero temperatures, where the injected state is mixed rather than pure, can be found in Refs.~\cite{Moskalets:2015ub,Moskalets:2017fh}.

\subsection{Two-particle emission}

For a source that emits two particles, we have~\cite{Grenier:2013gg}
\begin{eqnarray}
G ^{(1)}_{}\left(  t_{1}; t_{2} \right)
=
\frac{ e^{\frac{ i }{ \hbar } \mu \left(  t_{1} - t_{2} \right) } }{ v_{ \mu} }
\sum\limits_{ \alpha=1}^{2}
\eta_{ \alpha}
\psi_{ \alpha}^{*}\left( t_{1}  \right) \psi_{ \alpha}^{}\left( t_{2}  \right)
.
\label{05}
\end{eqnarray}
If the source injects two particles of the same kind, two electrons or two holes, we have $ \eta_{1} \eta_{2}=+1$, and the corresponding wave functions are orthogonal to each other, meaning that their overlap integral
\begin{eqnarray}
J = \int _{}^{ } dt \psi_{1}^{*}\left(  t \right) \psi_{2}\left(  t \right),
\label{06}
\end{eqnarray}
vanishes, $J_{+}=0$ (here the subscript denotes the product sign $\eta_{1} \eta_{2}$). This fact is the manifestation of the Pauli exclusion principle for fermions injected into the same quantum channel, see e.g. Ref.~\cite{Moskalets:2014ea}.

By contrast, if the two-particle source emits one electron and one hole, $ \eta_{1} \eta_{2} = -1$, the overlap integral is not necessarily zero, $J_{-} \ne 0$. To clarify the physics behind this observation, we need to analyse the two-particle wave function.

\subsection{Electron-hole emission and annihilation}

The fermionic two-particle wave function is represented by the following Slater determinant containing the single-particle wave functions, $\psi_{\alpha}$,
\begin{eqnarray}
\psi^{(2)}\left(  t_{1};t_{2} \right) &=&
\begin{vmatrix}
\psi_{1}\left(  t_{1} \right)
& \,\,\,
\psi_{2}\left(  t_{1} \right)
\\ \\
\psi_{1}\left(  t_{2} \right)
& \,\,\,
\psi_{2}\left(  t_{2} \right)
\end{vmatrix} .
\label{07}
\end{eqnarray}
The integral of the squared wave function gives us the number of injected particles,
\begin{eqnarray}
N &=&
\iint _{}^{ } dt_{1} dt_{2} \left | \psi^{(2)}\left(  t_{1};t_{2} \right)  \right |^{2}
= 2 \left(  1 - \left | J_{}  \right |^{2} \right) ,
\label{08}
\end{eqnarray}
having used the normalization of the single-particle wave functions according to Eq.~(\ref{03}) together with the definition of the overlap integral in Eq.~(\ref{06}).

Note that in the case of electron-hole injection, simply using $G ^{(1)}_{}$ for calculating the number of particles is incorrect, since $ \int\! dt G ^{(1)}_{}\left(  t;t \right)$ gives us the difference, not the sum of the number of injected particles.

If two particles of the same kind are injected, their wave functions are orthogonal, $J_{+} =0$, and the number of injected particles is $N_{+} = 2$, as expected. The injected state $\ket{\psi^{(2)}}$ is a two-particle state, $\ket{\psi^{(2)}} = \ket{ee}$ or $\ket{\psi^{(2)}} = \ket{hh}$, where $e$ stands for an electron and $h$ stands for a hole.

If an electron and a hole are injected, their wave functions are not orthogonal, $J_{-}  \ne 0$, and the number of injected particles is less than expected, $N_{-} < 2$, according to Eq.~(\ref{08}). One can interpret this suppression as being caused by the two injected particles annihilating each other with a probability given by the squared overlap integral, $\left | J_{-} \right |^{2}$.
One may think of the electron emitted by one source as being reabsorbed by the other source which emits a hole (or \emph{vice versa})~\cite{Splettstoesser:2008gc,Moskalets:2009dk,Moskalets:2013dl,Juergens:2011gu}. As such, the emitted state is a coherent superposition of the two-particle electron-hole state, $\ket{eh}$, and the state without any injected particles, the vacuum state $\ket{0}$,
\begin{eqnarray}
\ket{\psi^{(2)}} = \sqrt{1 - \left | J_{-} \right |^{2} } \ket{eh} + J_{-} \ket{0} .
\label{09}
\end{eqnarray}
As we now go on to show, the number of injected particles in Eq.~(\ref{08}) can be determined from shot noise measurements at low temperatures~\cite{Dubois:2013ul,Bocquillon:2012if}.

\subsection{Shot noise}

\begin{figure}[b]
\centering
\resizebox{0.75\columnwidth}{!}{\includegraphics{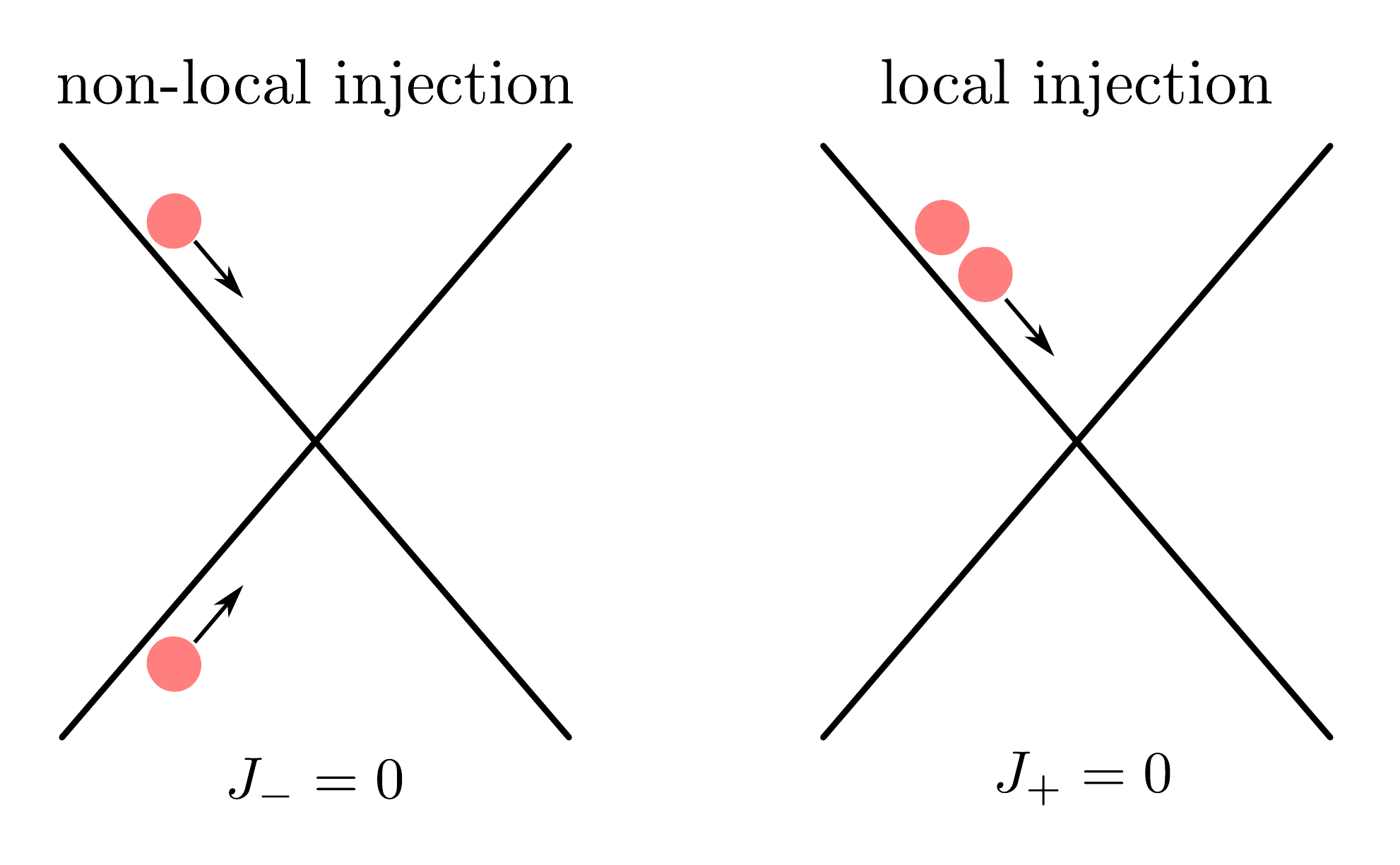} }
\caption{
Non-local and local injection of fermions on a beam splitter.
{\bf Left panel}:
If particles are injected into different input arms, the overlap integral in Eq.~(\ref{06}) describes the scattering on the beam splitter, subject to the Pauli exclusion principle for electron-electron or hole-hole interference, $J_{+} \ne 0$, but not for the simultaneous arrival of an electron and a hole, $J_{-} = 0$.
{\bf Right panel}:
If two particles are injected into the same input arm, the overlap integral rather characterizes the injection process.
The Pauli exclusion principle demands that the overlap vanishes, $J_{+}=0$, if two electrons or two holes are injected.
Otherwise, $J_{-} \ne 0$ is the amplitude for the particles to annihilate each other according to Eq.~(\ref{09}).
}
\label{fig2}
\end{figure}

Shot noise is generated if the stream of injected particles is partitioned on an electronic beam splitter as illustrated in Fig.~\ref{fig2}~\cite{Schottky:1918bo,Reznikov:1998kn,Blanter:2000wi,Buttiker:1992vr,Liu:1998wr}.
We first emit particles into one of the input arms using a periodically driven electron source, while the other input channel is grounded, as indicated in the right panel of Fig.~\ref{fig2}. In this case, the time-averaged cross-correlations of the outgoing currents, ${\cal P}_\text{out}$, is related to the excess first-order correlation function of the incoming excitations as~\cite{Moskalets:2016fm}
\begin{eqnarray}
\frac{{\cal P}_\text{out} }{ {\cal P}_{0}  } = -
\int\limits _{0}^{ {\cal T} _{0} }  dt
\int\limits _{- \infty}^{ \infty } d \tau  \left | v_{ \mu}
G^{(1)}_{ }(t_{}+ \tau;t_{} ) \right | ^{2}.
\label{10}
\end{eqnarray}
Here, the factor ${\cal P}_{0} = e^{2} R T / {\cal T} _{0}$ is given by the electron charge $e$, the reflection $R$ and transmission probabilities $T=1-R$ of the beam splitter, and the period of the drive ${\cal T} _{0}$. To keep the discussion simple, we assume that electrons injected in different periods do not overlap.
We can then extend the limits of the integral over $t$ to infinity. Upon substituting Eq.~(\ref{05}) into Eq.~(\ref{10}), combined with Eqs.~(\ref{03}) and (\ref{06}), we now arrive at an expression for the shot noise reading
\begin{eqnarray}
\frac{{\cal P}_\text{out} }{ {\cal P}_{0}  } &=& - N
= - 2 \left(  1 - \left | J \right |^{2} \right).
\label{11}
\end{eqnarray}
This result shows that the shot noise is given by the product of two factors: One, $-2  {\cal P}_{0}$, is the shot noise produced by two independent particles. The other one, $1 - \left | J \right |^{2}$, is the probability that the two particles do not annihilate each other.

\subsubsection{Annihilation versus the electronic Hong-Ou-Mandel effect}

Equation (\ref{11}) also describes the situation where electrons are injected into different input arms and interfere at the beam splitter as in a Hong-Ou-Mandel interferometer~\cite{Bocquillon:2013fp,Blanter:2000wi,Ferraro:2018um}. Note that we are discussing two-particle, not single-particle, interference in the following. Although the equation is formally the same for the two cases, the physical interpretation of the overlap integral in Eq. (\ref{06}) is completely different.

In the \textit{non-local} setup in the left panel of Fig.~\ref{fig2}, the two sources emit particles into different input channels, and the emitted electrons or holes ($ \eta_{1} \eta_{2} = +1$) are initially uncorrelated. They interfere when they arrive at the beam splitter, where the Pauli exclusion principle forces them to exit via different output arms~\cite{Blanter:2000wi}. In this case, the overlap of the wave functions at the beam splitter determines the suppression of the shot noise, and its exact value, $J_{+} \ne 0$, can be controlled by adjusting the relative emissions times. On the other hand, if an electron and a hole are injected, $ \eta_{1} \eta_{2} = -1$, they scatter on the beam splitter independently, since their energy is different such that the Pauli exclusion principle does not apply, and the overlap integral vanishes, $J_{-}=0$. We recall that we here consider zero temperature and refer the reader to Refs.~\cite{Jonckheere:2012cu,Rech:2016cd} for a discussion of electron-hole interference at nonzero temperatures.

By contrast, in the \textit{local} setup in the right panel of Fig.~\ref{fig2}, the two emitters are placed in the same input arm to make up a composite source. In this case, two injected electrons (or holes) with $ \eta_{1} \eta_{2} = +1$ interfere already upon emission. The Pauli exclusion principle forces the two injected electrons (or holes) to be in orthogonal quantum states and the overlap must vanish, $J_{+} = 0$, independently of the emission times. On the other hand, if one source emits an electron and the other a hole, $ \eta_{1} \eta_{2} = -1$, the overlap integral is not necessarily zero, $J_{-} \ne 0$ and it characterizes correlations between the entire injected state and the vacuum (the untouched Fermi sea) according to Eq.~(\ref{09}). More precisely, the overlap integral $J_{-}$ is just the quantum-mechanical amplitude of the annihilation process, in which an electron injected by one source is annihilated by a hole from the other source, such that no particles effectively are emitted towards the beam splitter. Importantly, the value of the overlap integral, $J_{-} \ne 0$, can in this case be continuously tuned by changing the emission times.

As we have seen, the shot noise suppression is clearly caused by different physical processes. In the non-local setup, the shot noise suppression occurs due to the interference of identical particles at the beam splitter~\cite{Olkhovskaya:2008en,Feve:2008im,Moskalets:2011jx}. By contrast, for the local setup, the shot noise suppression is due to the decreased number of injected particles, since electron and holes may annihilate each other close to the emitters. Shot noise suppression forms the basis of single-electron state tomography based on Hong-Ou-Mandel interferometry~\cite{Grenier:2011dv}, and one may anticipate that similar ideas can be developed based on the annihilation effect for the local setup. Importantly, with the sources placed close to each other as in the right panel of Fig.~\ref{fig2}, the setup becomes less susceptible to decoherence along the electron waveguide~\cite{Cabart:2018ef,Iyoda:2014cf}.

\section{Scattering theory of composite two-electron sources}
\label{s2}

We now describe the scattering theory of composite two-electron sources. To this end, we first consider a single-particle emitter that is attached to a chiral one-dimensional electron waveguide. The source is driven by a classical periodic field with period $ {\cal T} _{0} = 2 \pi/ \Omega$, for example an electric potential that is applied with a gate electrode.

The effect of a periodic driving field can be described by a Floquet scattering matrix $S_\text{F}$, whose elements $S_\text{F}\left(  E_{n}, E \right)$ are the photon-assisted amplitudes for an electron with energy $E$ to exchange $n$ energy quanta of size $ \hbar \Omega$ with the driving field such that its energy becomes $E_{n} = E + n \hbar \Omega$~\cite{Moskalets:2002hu}. The scattering amplitudes relate the electron operators in second quantization before and after the source, $ \hat a$ and $\hat b$, respectively, as
\begin{equation}
\hat b\left(  E_{n} \right) = S_\text{F}\left(  E_{n}, E \right) \hat a\left(  E \right)
\end{equation}
for annihilation operators and
\begin{equation}
\hat b^{ \dagger}\left(  E_{n} \right) = S_\text{F}^{*}\left(  E_{n}, E \right) \hat a^{ \dagger }\left(  E \right).
\end{equation}
for creation operators. The electrons before the source are in equilibrium and are characterized by the Fermi distribution function
\begin{equation}
f(E)  = \frac{1}{ 1 + \exp  \left[ \frac{ E - \mu }{ k _{B} \theta } \right]  },
\end{equation}
where $ \mu$ is Fermi energy, the temperature is $ \theta$, and $k _{B}$ is the Boltzmann constant. The quantum-mechanical average used in the definition of $G ^{(1)}_{}$ describes the fermionic operators before the source, where the particles are in equilibrium and we have
\begin{equation}
\left \langle \hat a^{ \dagger }\left(  E \right) \hat a^{ }\left(  E ^{\prime} \right) \right \rangle = \delta\left(  E - E ^{\prime} \right) f(E).
\end{equation}
In the wide band limit, the excess first-order correlation function then becomes~\cite{Haack:2013ch}
\begin{eqnarray}
G^{(1)}_{ }( t_{1};t_{2}) &=&
\frac{ 1 }{ h v_{ \mu} }
\int   d E    f\left( E \right)
e^{ \frac{ i }{ \hbar  } E   \left( t_{1} - t_{2} \right)  }
\left\{ S_\text{in}^{*}( t_{1}, E)  S_\text{in}( t_{2 }, E)    - 1 \right\} ,
\label{12}
\end{eqnarray}
where we have introduced the inverse Fourier transform of the Floquet amplitudes~\cite{Moskalets:2008ii}
\begin{eqnarray}
S_\text{in}\left(  t, E \right) = \sum\limits_{n=-\infty}^{\infty} e^{- i n \Omega t} S_\text{F}\left(  E_{n}, E  \right) .
\label{13}
\end{eqnarray}
These expressions allow us to describe dynamic single-electron emitters.

\subsection{Composite sources}

We can now describe two-particle sources composed of two single-particle emitters connected in series~\cite{Splettstoesser:2008gc}. If each source separately  operates as an ideal single-particle emitter, we can expect that the composite source will work as a two-particle emitter.

We place the two single-particle emitters, each described by the scattering matrices $S_\text{in}^\text{L}$ and $S_\text{in}^\text{R}$, in close proximity to form a composite source. To be specific, we choose $S_\text{in}^\text{L}$ to describe first emitter upstream. The total scattering amplitude of the composite source can then be calculated as~\cite{Moskalets:2013dl}
\begin{eqnarray}
S_\text{in} ^\text{tot}\left(  t, E \right) = \sum\limits_{n=-\infty}^{\infty} \int\limits _{0}^{ {\cal T} _{0} } \frac{ d \tau }{ {\cal T} _{0}  } e^{i n \Omega \left(  \tau - t \right)} S_\text{in} ^\text{R}\left(  t, E_{n} \right) S_\text{in} ^\text{L}\left(   \tau, E \right) ,
\label{14}
\end{eqnarray}
where we have neglected the distance between the sources. At zero temperature and in the wide band limit, the distance between the emitters can be accounted for simply by adjusting the injection times of the two emitters.

The correlation function of the composite source can be calculated using Eq.~(\ref{12}) by replacing $S_\text{in}$ with $S_\text{in}^\text{tot}$.
The correlation function then consists of three terms,
\begin{eqnarray}
G ^{(1)}_\text{tot} &=&  G ^{(1)}_\text{R} + G ^{(1)}_\text{L} + \delta G ^{(1)}_\text{LR},
\label{15}
\end{eqnarray}
where $G ^{(1)}_{ j}, j = \text{L, R},$ is given in Eq.~(\ref{12}) with $S_\text{in}$ replaced by $S_\text{in}^{j}$, while the last term accounts for the combined effect of the two single-particle emitters and reads
\begin{eqnarray}
\delta G^{(1)}_\text{LR} &=&
\frac{ 1 }{ h v_{ \mu} }
\int   d E    f\left( E \right)
e^{ \frac{ i }{ \hbar  } E   \left( t_{1} - t_{2} \right)  }
\sum\limits_{m,n}^{}
\iint\limits _{0}^{ {\cal T} _{0} }
\frac{ d \tau ^{\prime} }{ {\cal T} _{0}  }
\frac{ d \tau }{ {\cal T} _{0}  }
e^{-i m \Omega \left(  \tau ^{\prime} - t_{1} \right)}
e^{i n \Omega \left(  \tau - t_{2} \right)}
\label{16} \\
&&
\times
\left\{
\left[ S_\text{in} ^\text{R}\left(  t_{1}, E_{m} \right)  \right]^{*}
S_\text{in} ^\text{R}\left(  t_{2}, E_{n} \right)
-1
\right\}
\left\{
\left[ S_\text{in} ^\text{L}\left(  \tau ^{\prime}, E_{} \right)  \right]^{*}
S_\text{in} ^\text{L}\left(  \tau, E_{} \right)
-1
\right\}
.
\nonumber
\end{eqnarray}
With these expressions, we can now describe the composite two-particle sources.

\section{Examples}
\label{s3}

We are now ready to discuss examples of composite two-particle sources. We restrict ourselves to zero temperature, $ \theta = 0$, where the Fermi distribution is a step function, $f\left(  E \right) = \Theta\left(  \mu - E \right)$, with $ \Theta\left(  x \right)$ being the Heaviside function.

\subsection{Adiabatic injection}

If the scattering amplitude of each single-particle source, $S_\text{in}^{j}$, is energy-independent, the total amplitude is simply given by the product, $S_\text{in}^\text{tot} \left(  t \right) = S_\text{in}^\text{R}\left(  t \right) S_\text{in}^\text{L}\left(  t \right)$~\cite{Splettstoesser:2008gc,Moskalets:2014ea}. The expression for the correlation function also greatly simplifies. In particular, at zero temperature, Eqs.~(\ref{12}), (\ref{15}), and (\ref{16}) in combination give us
\begin{subequations}\label{17}
\begin{eqnarray}
G ^{(1)}_{j}\left(  t_{1}; t_{2} \right) &=&
\frac{ e ^{\frac{ i }{ \hbar } \mu \left(  t_{1} - t_{2} \right) } }{ v_{ \mu} }
\frac{
	\left[ S_\text{in} ^{j}\left(  t_{1} \right)  \right]^{*}
	S_\text{in} ^{j}\left(  t_{2} \right) - 1
}{ 2 \pi i \left(  t_{1} - t_{2} \right)  } ,
\\
G ^{(1)}_\text{tot}\left(  t_{1}; t_{2} \right) &=&
G ^{(1)}_\text{R}\left(  t_{1}; t_{2} \right)
+
\left[ S_\text{in} ^\text{R}\left(  t_{1} \right)  \right]^{*}
G ^{(1)}_\text{L}\left(  t_{1}; t_{2} \right)
S_\text{in} ^\text{R}\left(  t_{2} \right)
.
\end{eqnarray}
\end{subequations}
The last equation admits the following interpretation:
the quantum state injected by the composite source is the sum of the quantum state injected by the downstream source and the state injected by the upstream source after being modified by the downstream one. Although this might be a rather artificial interpretation, since we can't really divide the quantum phase-coherent system into parts, it provides a useful description of what is happening in the system.

Equation (\ref{17}) tells us that in the adiabatic case, if each source separately works as a single-electron emitter, the two emitters form a perfect two-particle source which excites no unwanted electron-hole pairs. Indeed, if the correlation function of each individual source has the form given in Eq.~(\ref{02}), $G ^{(1)}_{j}\left(  t_{1}; t_{2} \right) \sim  \psi_{j}^{*}\left(  t_{1} \right) \psi_{j}^{}\left(  t_{2} \right)$, $j=\text{L, R}$, the correlation function of the composite source, $G ^{(1)}_\text{tot}$, is given by Eq.~(\ref{05}) with, for instance, $ \psi_{1}\left(  t \right) = \psi_\text{R}\left(  t \right)$ and $ \psi_{2}\left(  t \right) = S_\text{in}^\text{R}\left(  t \right) \psi_\text{L}\left(  t \right)$.
We note that in the adiabatic limit, the scattering amplitude of an emitter coupled to a chiral one-dimensional channel is just a phase factor, $S_\text{in}^{j} = e^{i \sigma_{j}}$, so that the modulus is unity, $| S_\text{in}^{j} |=1$.
For this reason, $ \psi_{2}$ is normalized if $ \psi_\text{L}$ is also normalized. This line of reasoning can be generalized to any number of sources operating in the adiabatic regime~\cite{Moskalets:2015kx}.

\subsection{Levitons}

A paradigmatic example of an energy independent source is the case of an AC voltage applied to a metallic contact from which a chiral edge state originates.
The scattering amplitude is the phase factor, $S _\text{in} ^{}(t) = \exp [ - i \left( e/ \hbar \right)  \int_{- \infty }^{ t } dt ^{\prime} V ^{}(t ^{\prime})  ]$, where $V(t)$ is the time-dependent voltage~\cite{Grenier:2013gg}.
If the voltage is a sequence of Lorentzian voltage pulses of a definite amplitude~\cite{Dubois:2013fs,Gaury:2014jz,Dasenbrook:2013tt,Belzig:2016jz,Rech:2017be,Suzuki:2017er,Ronetti:2017vd,Safi:2019dq,Burset:2019ha,Dashti:2019ts},
\begin{eqnarray}
eV(t) = \eta \sum\limits_{m=-\infty}^{\infty} \frac{2 \hbar  \Gamma _{\tau} }{\left( t - \tau - m {\cal T} _{0} \right)^{2} +  \Gamma _{\tau}^{2} },
\label{L01}
\end{eqnarray}
a sequence of single electrons ($ \eta = +1$) or holes ($ \eta = -1$) are injected~\cite{Levitov:1996ie,Ivanov:1997wz,Keeling:2006hq}.
These particles are called {\it levitons}~\cite{Dubois:2013ul}. The parameter $ \Gamma _{\tau}$ is the half-width of the density profile of the wave packets.
It also defines the energy of each leviton, $ {\cal E} = \hbar/ (  2 \Gamma _{\tau} )$~\cite{Keeling:2006hq}.
The parameter $ \tau$ determines the position of the peak of the wave packet within the period, $0 < \tau < {\cal T} _{0}$.
In general, when $ {\cal T} _{0} \sim \Gamma _{\tau}$, the particles injected during different periods are overlapping, the resulting state is thus strictly speaking a multi-particle state, and the corresponding wave functions were analysed in Ref.~\cite{Moskalets:2015kx}.

To simplify our analysis of two Lorentzian voltage pulses being applied per period, we consider the limiting case $ {\cal T} _{0} \gg \Gamma _{\tau}$ and restrict ourselves to a single long period. This approach can also be employed when Floquet scattering theory is used to describe a non-periodic process as we will see in Sec.~\ref{exc}~\cite{Moskalets:2017fh}. The applied voltage is then the sum of two Lorentzian pulses, $V_\text{tot} = V_\text{L}$ + $V_\text{R}$, with
\begin{eqnarray}
eV_{j}(t) = \eta_{j} \frac{2 \hbar  \Gamma _{j} }{\left( t - \tau_{j}  \right)^{2} +  \Gamma _{j}^{2} }.
\label{19}
\end{eqnarray}
Here, the time $t$ extends over the full, long perid $ {\cal T} _{0} \gg \Gamma _{j}$. The voltages $V_\text{L,R}$ now play the role of single-particle sources, while the total voltage $V_\text{tot}$ effectively is our composite two-particle source. The corresponding scattering amplitudes read~\cite{Keeling:2008ft}
\begin{eqnarray}
S_\text{in}^{j}\left(  t \right) =  \frac{ t - \tau_{j} + i \eta_{j}\Gamma _{j}  }{  t - \tau_{j} - i \eta_{j}\Gamma _{j}  }  .
\label{20}
\end{eqnarray}
According to Eq.~(\ref{17}), the correlation function of the composite source  has the form of Eq.~(\ref{05}) with $ \eta_{1} = \eta_\text{R}$, $ \eta_{2} = \eta_\text{L}$ and the following wave functions~\cite{Glattli:2016tr,Grenier:2013gg}
\begin{equation}
\psi_{1}\left(  t \right) = \frac{ \sqrt{ \Gamma _\text{R}/ \pi } }{ t - \tau_\text{R} - i \eta_\text{R} \Gamma _\text{R}  } ,
\, \quad \,
\psi_{2}\left(  t \right) = \frac{ \sqrt{ \Gamma _\text{L}/ \pi } }{ t - \tau_\text{L} - i \eta_\text{L} \Gamma _\text{L}  }
\frac{ t - \tau_\text{R} + i \eta_\text{R}\Gamma _\text{R}  }{  t - \tau_\text{R} - i \eta_\text{R}\Gamma _\text{R}  }  ,
\label{21}
\end{equation}
which are normalized according to Eq.~(\ref{03}).
When both voltages have the same sign, $ \eta_\text{L} = \eta_\text{R}$, the overlap integral in Eq.~(\ref{06}) is zero, $J_{+}=0$, recalling that the subscript stands for the sign of the product $ \eta_\text{L} \eta_\text{R}$.
In other words, when both voltages excite particles with the same charge (two electrons or two holes), the particles occupy orthogonal states in accordance with the Pauli exclusion principle.

By contrast, if the two voltages $V_\text{L}$ and $V_\text{R}$ excite particles of opposite charge, $ \eta_\text{L} = - \eta_\text{R}$, the overlap integral is non-zero, and one may interpret this as the particle generated by one source being annihilated by the other with probability $\left | J_{-} \right |^{2}$. Indeed, using Eqs.~(\ref{06}), (\ref{08}), and (\ref{21}), the number of emitted particles reads
\begin{equation}\label{22}
N =
2\left( 1 - \left | J_{-} \right |^{2} \right) \leq 2,
\, \quad \,
\left | J_{-} \right |^{2} =
\frac{ 4 \Gamma _\text{L}^{} \Gamma _\text{R}^{} }{ \tau_{}^{2} + \left(  \Gamma _\text{L}^{} + \Gamma _\text{R}^{} \right)^{2}  } ,
\end{equation}
where $ \tau = \tau_\text{L} - \tau_\text{R}$ is the time difference between the two voltage pulses. If the electron and hole wave packets have the same width, $ \Gamma _\text{L} = \Gamma _\text{R}$, and are injected simultaneously, $ \tau =0$, the annihilation is complete and no particles are emitted, $N=0$~\cite{Splettstoesser:2008gc}. For this reason, we can refer to a hole-like leviton as an {\it anti-leviton}, and we see how a leviton created by one voltage pulse and a hole created by another voltage pulse can completely annihilate each other. Of course, in this simple case, the voltage pulses simply cancel each other, such that the total applied voltage vanishes.

Interestingly, the overlap integral in Eq.~(\ref{22}) has the same time dependence as the wave function squared in Eq.~(\ref{21}), however, with both the amplitude and the width of the wave packet being renormalized.
This fact can be used to access the time-dependent density profile of a single-electron wave packet via a time-averaging measurement like the shot noise measurement described by Eq.~(\ref{11}).

\subsection{Non-adiabatic injection}
\label{exc}

We now move on to non-adiabatic particle injection. Theory predicts that an electron and a hole injected in an irreversible decay process do not annihilate each other~\cite{Moskalets:2013dl}. Therefore, one may speculate that complete annihilation only occurs in the adiabatic regime. To support this expectation, we now discuss a single-particle source that can be tuned from the adiabatic to the non-adiabatic regime. Specifically, we consider single-particle emission into a Fermi sea from a single quantum level that can be varied linearly in time with rapidity $c$. The corresponding scattering amplitude reads~\cite{Keeling:2008ft}
\begin{equation}
S_\text{in} ^{}(t, E) =
1-  2
\int\limits _{0}^{ \infty } d \xi_{}
e^{ -  \xi_{}  }
e^{-i  \frac{  t_{} - \tau }{ \Gamma _{\tau}  }  \xi_{} }
e^{ i   \zeta_{}  \xi_{}^{2} }
e^{ i \zeta_{} \frac{ E - \mu }{ {\cal E}  }   \xi_{} },
\label{23}
\end{equation}
where $ \tau$ is the time when the quantum level would cross the Fermi energy $ \mu$, if there were no coupling. The width of the injected wave packet due to the coupling to the Fermi sea is denoted by $\Gamma _{\tau}$. This width determines the time window during which the broadened quantum-dot level crosses the Fermi level. The coupling strength can also be characterized by the dwell time, $ \tau_{D}$, which is the time it takes an electron to decay from the quantum level at a fixed energy above the Fermi level. In addition, the crossing time $ \Gamma _{\tau}$ depends on the rapidity $c$. Thus, the parameter that enters Eq.~(\ref{23}) is defined as $ \Gamma _{\tau} = \hbar / (  2 c_{} \tau_{D} )$.
Moreover, the parameter $\zeta$ is given by the ratio of the two characteristic time scales, $\zeta = \tau_{D}/ \Gamma _{\tau}$, and it controls how adiabatic the driving is. Specifically, with $ \zeta=0$, we recover the adiabatic solution in Eq.~(\ref{20}). Finally,
$ {\cal E} = \hbar/ (  2 \Gamma _{\tau} )$ is the energy of the injected particle relative to the Fermi level~\cite{Moskalets:2016fm}.

The wave function of the injected particle reads $ \Psi(t) = e^{ - \frac{ i }{ \hbar  } \mu  t  } \psi(t)$ with~\cite{Keeling:2008ft}
\begin{eqnarray}
\psi_{} \left( t \right) &=&
\frac{  1 }{  \sqrt{\pi  \left | \Gamma _{\tau} \right | } }
\int\limits_{0}^{ \infty} d \xi
e^{ -   \xi  }
e^{  - i  \frac{ t - \tau }{ \Gamma _{ \tau}  }   \xi }
e^{ i  \zeta_{} \xi^{2} } .
\label{24}
\end{eqnarray}
The wave function is normalized according to Eq.~(\ref{03}), and we note that the two parameters $ \Gamma _{\tau}$ and $ \zeta$ both depend on the rapidity $c$. If the rapidity is positive (negative), $c>0$ ($c<0$), an electron (hole) is injected.
To take this into account, we use $ \eta \Gamma _{\tau}$ and $ \eta \zeta$, with both $ \Gamma _{\tau}$ and $ \zeta$ being positive, and $ \eta = 1 (-1)$ for electron (hole) injection.

\subsubsection{Two-particle injection}

We now take two quantum dot levels as above and attach them in series to a chiral waveguide. As before, we use the subscript L for the upstream source and R for the downstream source. The scattering amplitude of each source reads
\begin{eqnarray}
S_\text{in} ^{j}(t, E) &=&
1-  2
\int\limits _{0}^{ \infty } d \xi_{}
e^{ -  \xi_{}  }
e^{-i \eta_{j}  \frac{  t_{} - \tau_{j} }{ \Gamma _{j}  }  \xi_{} }
e^{ i \eta_{j}   \zeta_{j}  \xi_{}^{2} }
e^{ i \zeta_{j}  \frac{ E - \mu }{ {\cal E}_{j}  }  \xi_{} } ,\,\,\, j = \text{L, R},
\label{25}
\end{eqnarray}
where the last exponential factor does not contain $ \eta_{j}$, since the ratio $ \zeta_{j}/ {\cal E}_{j}$ is the same for electron ($\eta_{j} = 1$) and hole ($\eta_{j} = -1$) injection.

Based on this expression, we can calculate the scattering amplitude of the composite source $S_\text{in}^\text{tot}$ using Eq.~(\ref{14}) and, furthermore, the correlation function $G ^{(1)}_\text{tot}$ using Eq.~(\ref{12}). The calculation is in principle straightforward, however, it is rather lengthy, and the details are presented in Appendix~\ref{appA} of the Supplementary Material.

At zero temperature, the result has the form of Eq.~(\ref{05}) with $ \eta_{1} = \eta_\text{R}$ and $ \psi_{1}(t) = \psi_\text{R}(t)$, where $\psi_\text{R}$ is the wave function of a particle that would be injected by the downstream source on its own,
\begin{eqnarray}
\psi_\text{R} \left( t \right) &=&
\frac{  1 }{  \sqrt{\pi   \Gamma _\text{R}  } }
\int\limits_{0}^{ \infty} d \xi
e^{ -   \xi  }
e^{  - i \eta_\text{R} \frac{ t - \tau_\text{R} }{ \Gamma _{ R}  }   \xi }
e^{ i \eta_\text{R} \zeta_\text{R} \xi^{2} } .
\label{26}
\end{eqnarray}
For the second contribution in Eq.~(\ref{05}), we have $ \eta_{2} = \eta_\text{L}$ and the wave function
\begin{eqnarray}
\psi_{2} \left( t \right) &=&
\frac{  1 }{  \sqrt{\pi  \Gamma _\text{L} } }
\int\limits_{0}^{ \infty} d \xi
e^{ -   \xi_{}  }
e^{  - i \eta_\text{L}  \frac{ t - \tau_\text{L} }{ \Gamma _\text{L}  }   \xi }
e^{ i \eta_\text{L}  \zeta_\text{L} \xi^{2} }
\nonumber \\
&&
\times
\left\{ 1 -2
\int\limits _{0}^{ \infty } d \chi
e^{ - \chi_{}  }
e^{-i \eta_\text{R}  \frac{  t_{} - \tau_\text{R}}{ \Gamma _\text{R}  }  \chi }
e^{ i \eta_\text{R}  \zeta_\text{R}  \chi^{2} }
e^{i \eta_\text{L}   \zeta_\text{R}  \frac{  \Gamma _\text{R} }{  \Gamma _\text{L}   } 2 \xi_{} \chi }  \right\} .
\label{27}
\end{eqnarray}
The wave function is normalized as shown in Appendix~\ref{appB} of the Supplementary Material.
Note that the equation above consists of two distinct terms, $ \psi_{2} = \psi_\text{L} + \delta \psi_\text{LR}$,  as compared to Eqs.~(\ref{15}) and (\ref{16}). First, $ \psi_\text{L}$ is given by Eq.~(\ref{26}) with the subscript R replaced by L, i.e., it represents a particle injected upstream in the absence of a downstream source. Second, $\delta \psi_\text{LR}$ describes the joint effect of the two sources.
The term $\delta \psi_\text{LR}$ is important, because it is responsible for the orthogonalization of $ \psi_{1}$ and $ \psi_{2}$ if $ \eta_\text{L} \eta_\text{R} =+1$ and for the annihilation effect if $ \eta_\text{L} \eta_\text{R} =-1$.

It is easy to show that the two wave functions are orthogonal, if $\eta_\text{L} = \eta_\text{R}$. In other words, the overlap integral in Eq.~(\ref{06}) vanishes, $J_{+} = 0$, with the subscript referring to the sign of the product $ \eta_\text{L} \eta_\text{R}$. The details of these calculations are presented in Appendix~\ref{appC} of the Supplementary Material. Hence, if the sources emit particles of the same kind, electrons or holes, the resulting two-particle state contains particles in orthogonal states as required by the Pauli exclusion principle. By contrast, if $ \eta_\text{L} = - \eta_\text{R}$, the two wave functions are no longer orthogonal (see Appendix~\ref{appD} of the Supplementary Material for further details), and we have
\begin{equation}
J_{-}\left(  \bar t \right) =
\frac{ 2 \sqrt{ \Gamma _\text{R} \Gamma _\text{L} } }{ \Gamma _\text{R} + \Gamma _\text{L}  }
\int\limits _{ 0}^{ \infty } d \xi
e^{ -  \xi }
e^{i  \bar t \xi    }
e^{ i  \bar \zeta \xi^{2}  }
, \,
\bar t = \frac{ \eta_\text{R} \tau_\text{R} + \eta_\text{L} \tau_\text{L} }{ \Gamma _\text{R} + \Gamma _\text{L}  }
, \,
\bar \zeta = \eta_\text{L}  \frac{ \Gamma _\text{R}^{2} \zeta_\text{R} + \Gamma _\text{L}^{2} \zeta_\text{L} }{ \left(  \Gamma _\text{R} + \Gamma _\text{L} \right)^{2} }
.
\label{28}
\end{equation}

As already discussed, a nonzero overlap integral, $J_{-} \ne 0$, indicates that the particles partially annihilate each other according to Eqs.~(\ref{08}) and (\ref{22}), and the degree of annihilation, $\left | J_{-} \right |^{2}$, can be accessed through shot noise measurements as shown in Eq.~(\ref{11}).
We note that for adiabatic injection, $ \zeta_\text{L} = \zeta_\text{R}$, Eq.~(\ref{28}) agrees with Eq.~(\ref{22}).

Similar to the case of leviton injection in Eq.~(\ref{22}) and the subsequent discussion, the overlap integral $J_{-}\left(  - \bar t \right)$ in Eq.~(\ref{28}) has the form of the wave function of a single emitter, but with the renormalized nonadiabaticity parameter $ \zeta \to \bar \zeta$, cf. $ \psi\left(  \left[ t- \tau \right]/ \Gamma _{\tau} \right)$ in Eq.~(\ref{24}).
Therefore, for a composite source consisting of two identical emitters, one can experimentally access the time-dependent electron density profile injected by a single-particle source by measuring the probability of annihilation as a function of the time delay between the emitters.

\begin{figure}
\centering
\resizebox{0.85\columnwidth}{!}{\includegraphics{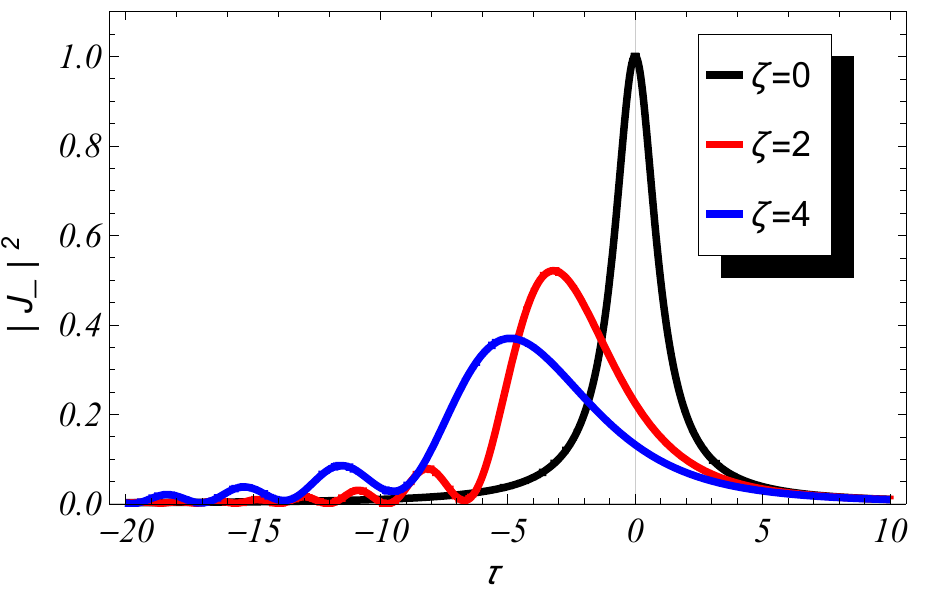} }
\caption{
The overlap integral $\left | J_{-} \right |^{2}$ in Eq.~(\ref{28}) as a function of the scaled time delay $ \tau \equiv \bar t$ for the renormalized non-adiabaticity parameter $ \zeta \equiv \bar \zeta = 0, 2, 4$.
The overlap integral is given in units of $2 \sqrt{ \Gamma _\text{R} \Gamma _\text{L} } / \left(  \Gamma _\text{R} + \Gamma _\text{L} \right)$.
}
\label{fig3}
\end{figure}

In Fig.~\ref{fig3}, we show the overlap $\left | J_{-} \right |^{2}$ as a function of the time delay between particle emissions for different values of the non-adiabaticity parameter. The maximum annihilation takes place in the case of adiabatic injection, $ \zeta_\text{L} = \zeta_\text{R} = 0$ (black line). In this case, a zero time delay ($ \tau =0$) leads to the transfer of an electron from one source to the other. Beyond the adiabatic regime, an increase in the rapidity suppresses the degree of annihilation at the maximum, and also it increases the asymmetry of the peaks. This behavior is in full agreement with a similar modification of the single-electron wave functions found in Ref.~\cite{Keeling:2008ft}.
The wavy structure on the left slope is a manifestation of quantum-mechanical interference {\it in time} that occurs during tunneling from a quantum level, whose position changes with time~\cite{Moskalets:2017fh}. An experimental observation of such a delicate quantum effect would be an important step towards the development of a time-resolved detector of single- and few-electron quantum states.

\section{Conclusions}

We have presented a Floquet scattering  theory of composite two-particle sources composed of several single-particle emitters connected in series to a chiral waveguide. The setup can include more than two single-particle emitters, and the individual emitters may be of different types.
For example, the combination of a leviton source with a quantum capacitor is within experimental reach, and it would enable the use of one source to characterize the other~\cite{Burset:2019ha}.

Using our theory, we have analysed several situations where ideal two-particle injection can be achieved. In particular, we have considered adiabatic injection using emitters with an energy-independent scattering amplitude. If both single-particle emitters operate under ideal conditions, the composite source emits exactly two particles without exciting any unwanted electron-hole pairs. An example of such an emitter is provided by a leviton source.

Going beyond the adiabatic regime, we have analysed a setup with two quantum levels that are shifted with a constant rapidity across the Fermi level of the external reservoir. This setup enables perfect two-particle injection both for adiabatic and non-adiabatic working conditions, where the injected wave packets have symmetric and asymmetric density profiles, respectively.

As an interesting new application of composite two-particle sources, we have examined the regime where one source injects an electron and the other a hole.
In this case, the electron and the hole may annihilate each other, effectively meaning that one source reabsorbs the particle that was emitted by the other.
If the probability of annihilation is less than one, a superposition of a two-fermion state and the vacuum state is formed. The amplitude of the annihilation process is given by the overlap of the injected single-particle quantum states. This fact can be used to develop single-electron tomography protocols based on this reabsorption effect. The advantage of our approach lies in the compactness of the setup, which helps avoiding decoherence, which is a major problem for quantum coherent electronics.

\begin{acknowledgement}
M.~M.~acknowledges the warm hospitality of Aalto University, support from the Aalto Science Institute through its Visiting Fellow Programme, and support from the Ministry of Education and Science of Ukraine (project No. 0119U002565).
P.~B.~acknowledges support from the European Union's Horizon 2020 research and innovation program under the Marie Sk\l odowska-Curie Grant No. 743884. The work was supported by Academy of Finland (projects No. 308515 and 312299).
\end{acknowledgement}

\newpage
\setcounter{page}{1}
\numberwithin{equation}{section}

\appendix

\noindent
{\LARGE
{\rm Supplementary Material for}\\
\ \\
{\bf Composite two-particle sources} \\
\ \\
}
\noindent
\author{Michael Moskalets\inst{1},
Janne Kotilahti\inst{2},
Pablo Burset\inst{2},
and Christian Flindt\inst{2}
}

{\small
\vspace{2mm}
\noindent
\author\inst{1}{Department of Metal and Semiconductor Physics, NTU ``Kharkiv Polytechnic Institute", 61002 Kharkiv, Ukraine}

\vspace{1mm}
\noindent
\author\inst{2}{Department of Applied Physics, Aalto University, 00076 Aalto, Finland}

\vspace{2mm}
\noindent
\thanks{\email{michael.moskalets@gmail.com}}
}

\section{Two-particle source composed of two quantum levels raising at a constant rapidity}
\label{appA}

The scattering amplitude of the composite source, $S_\text{in}^\text{tot}$, is calculated using Eq.~(\ref{14}) of the main text with $S_\text{in}^{j}$ being the scattering amplitude of a single source $j=\text{L, R}$.
Recall that the source with a scattering amplitude $S_\text{in}^\text{L}$ is located  upstream the source with a scattering amplitude  $S_\text{in}^\text{R}$.

For definiteness, we consider both sources to be in a single-electron injection regime, which is described by the scattering amplitude given in Eq.~(\ref{25}) of the main text with $ \eta_\text{L} = \eta_\text{R} = +1$,
\begin{subequations}\label{a01}
\begin{eqnarray}
S_\text{in} ^\text{R}(t, \epsilon) &=&
1-
\int\limits _{0}^{ \infty } d \xi_\text{R}
e^{ - \frac{ \xi_\text{R} }{2 } }
e^{-i \xi_\text{R} \frac{  t_{} }{ 2 \Gamma _\text{R}  }   }
e^{ i \xi_\text{R}^{2}  \frac{\zeta_\text{R}  }{ 4}  }
e^{ i \frac{ \epsilon_{} }{ {\cal E}_\text{R}  } \xi_\text{R} \frac{\zeta_\text{R} }{ 2 } } ,
\\
S_\text{in} ^\text{L}(t, \epsilon) &=&
1-
\int\limits _{0}^{ \infty } d \xi_\text{L}
e^{ - \frac{ \xi_\text{L} }{2 } }
e^{-i \xi_\text{L} \frac{  t_{} - \tau_\text{L} }{ 2 \Gamma _\text{L}  }   }
e^{ i \xi_\text{L}^{2}  \frac{\zeta_\text{L}  }{ 4}  }
e^{ i \frac{ \epsilon_{} }{ {\cal E}_\text{L}  } \xi_\text{L} \frac{\zeta_\text{L} }{ 2 } } .
\end{eqnarray}
\end{subequations}
Note that the notation here is slightly different from Eq.~(\ref{25}) of the main text.
First, we introduced $ \epsilon = E - \mu$, and second, the integration variable is doubled, $ 2\xi \to \xi_{j}$.
Without loss of generality, here we set $ \tau_\text{R} =0$.

The total correlation function, $G ^{(1)}_\text{tot} =  G ^{(1)}_\text{R} + G ^{(1)}_\text{L} + \delta G ^{(1)}_\text{LR}$, Eq.~(\ref{15}) of the main text, is represented as the sum of three terms.
The first two, $G ^{(1)}_{j}$ ($j = \text{L, R}$) with
\begin{subequations}\label{a02}
\begin{eqnarray}
G ^{(1)}_{j}\left(  t_{1}; t_{2} \right) &=& \frac{ e^{  \frac{ i }{ \hbar  } \mu  \left(  t_{1} - t_{2} \right)  } }{ v_{ \mu}  } \psi_{j}^{*} \left(  t_{1} \right)\psi_{j}\left(  t_{2} \right) ,
\\
\psi_{j} \left( t \right) &=&
\frac{  1 }{  \sqrt{\pi  \Gamma _{j} } }
\int\limits_{0}^{ \infty} d \xi
e^{ -   \xi  }
e^{  - i \xi \frac{ t - \tau_{j} }{ \Gamma _{j}  }    }
e^{ i \xi^{2} \zeta_{j}  } ,
\end{eqnarray}
\end{subequations}
are the contribution due to each source working independently.
The additional contribution $ \delta G ^{(1)}_\text{LR}$, Eq.~(\ref{16}) of the main text, is due to the joint work of both sources.

To calculate the latter term we proceed as follows.
First, we consider the two factors, $P_\text{L} = -1 + S_\text{in} ^{\text{L}*}\left(  \tau ^{\prime}, E_{} \right) S_\text{in} ^\text{L}\left(  \tau, E_{} \right)$ and $P_\text{R} = -1 + S_\text{in} ^{\text{R}*}\left(  t_{1}, E_{m} \right) S_\text{in} ^\text{R}\left(  t_{2}, E_{n} \right)$, separately and simplify them.
Recall that $E_{n} = E + n \hbar \Omega$.

\subsection{The factor $P_\text{L}$}

We represent this factor as the sum of three terms, $P_\text{L} =  A_\text{L} + B_\text{L} + C_\text{L}$, with
\begin{eqnarray}
A_\text{L} &=&
-
\int\limits _{0}^{ \infty } d \xi_\text{L} ^{\prime}
e^{ - \frac{ \xi_\text{L} ^{\prime} }{2 } }
e^{i \xi_\text{L}^{\prime}  \frac{ \tau ^{\prime} - \tau_\text{L} }{ 2 \Gamma _\text{L}  }   }
e^{- i \left(  \xi_\text{L}^{\prime} \right)^{2}  \frac{\zeta_\text{L}  }{ 4}  }
e^{ -i \frac{ \epsilon_{} }{ {\cal E}_\text{L}  } \xi_\text{L}^{\prime} \frac{\zeta_\text{L} }{ 2 } } ,
\label{a03}  \\
B_\text{L} &=&
-
\int\limits _{0}^{ \infty } d \xi_\text{L}
e^{ - \frac{ \xi_\text{L} }{2 } }
e^{-i \xi_\text{L} \frac{  \tau - \tau_\text{L}  }{ 2 \Gamma _\text{L}  }   }
e^{ i \xi_\text{L}^{2}  \frac{\zeta_\text{L}  }{ 4}  }
e^{ i \frac{ \epsilon_{} }{ {\cal E}_\text{L}  }  \xi_\text{L} \frac{\zeta_\text{L} }{ 2 } } ,
\nonumber \\
C_\text{L} &=&
\int\limits _{0}^{ \infty } d \xi_\text{L} ^{\prime}
e^{ - \frac{ \xi_\text{L} ^{\prime} }{2 } }
e^{i \xi_\text{L}^{\prime}  \frac{ \tau ^{\prime} - \tau_\text{L} }{ 2 \Gamma _\text{L}  }   }
e^{- i \left(  \xi_\text{L}^{\prime} \right)^{2}  \frac{\zeta_\text{L}  }{ 4}  }
e^{ -i \frac{ \epsilon_{} }{ {\cal E}_\text{L}  }  \xi_\text{L}^{\prime} \frac{\zeta_\text{L} }{ 2 } }
\int\limits _{0}^{ \infty } d \xi_\text{L}
e^{ - \frac{ \xi_\text{L} }{2 } }
e^{-i \xi_\text{L} \frac{  \tau - \tau_\text{L}  }{ 2 \Gamma _\text{L}  }   }
e^{ i \xi_\text{L}^{2}  \frac{\zeta_\text{L}  }{ 4}  }
e^{ i \frac{ \epsilon_{} }{ {\cal E}_\text{L}  }  \xi_\text{L} \frac{\zeta_\text{L} }{ 2 } } .
\nonumber
\end{eqnarray}
Then we split the third term as $C_\text{L} = D_\text{L} + E_\text{L}$, with
\begin{eqnarray}
D_\text{L} &=&
\int\limits _{0}^{ \infty } d a
e^{ - \frac{ a }{2 } }
e^{i a \frac{ \tau ^{\prime} - \tau_\text{L} }{ 2 \Gamma _\text{L}  }   }
e^{- i a^{2}  \frac{\zeta_\text{L}  }{ 4}  }
e^{ -i \frac{ \epsilon_{} }{ {\cal E}_\text{L}  }  a \frac{\zeta_\text{L} }{ 2 } }
\int\limits _{0}^{ \infty } d \xi_\text{L}
e^{ - \xi_\text{L}  }
e^{-i \xi_\text{L} \frac{  \tau - \tau ^{\prime} }{ 2 \Gamma _\text{L}  }   }
e^{ -i a \xi_\text{L}  \frac{\zeta_\text{L}  }{ 2}  }  ,
\nonumber \\
E_\text{L} &=&
\int\limits _{0}^{ \infty } d a
e^{ - \frac{ a }{2 } }
e^{-i a \frac{  \tau - \tau_\text{L}  }{ 2 \Gamma _\text{L}  }   }
e^{ i a^{2}  \frac{\zeta_\text{L}  }{ 4}  }
e^{ i \frac{ \epsilon_{} }{ {\cal E}_\text{L}  }  a \frac{\zeta_\text{L} }{ 2 } }
\int\limits _{0}^{ \infty } d \xi_\text{L} ^{\prime}
e^{ - \xi_\text{L} ^{\prime}  }
e^{i \xi_\text{L}^{\prime}  \frac{ \tau^{\prime} - \tau  }{ 2 \Gamma _\text{L}  }   }
e^{ i a \xi_\text{L} ^{\prime}  \frac{\zeta_\text{L}  }{ 2}  }  .
\label{a04}
\end{eqnarray}
In these terms we introduced $a>0$ instead of $ \xi_\text{L} ^{\prime} = \xi_\text{L} + a$ for $D_\text{L}$ and instead of $ \xi_\text{L}  = \xi_\text{L}^{\prime} + a$ for $E_\text{L}$.
Then, we rearrange the four terms, $A_\text{L}, B_\text{L}, D_\text{L}$ and $E_\text{L}$, as follows, $P_\text{L} = F_\text{L} + G_\text{L}$, where
\begin{eqnarray}
F_\text{L} = D_\text{L} +A_\text{L} &=&
\int\limits _{0}^{ \infty } d a
e^{ - \frac{ a }{2 } }
e^{i a \frac{ \tau ^{\prime} - \tau_\text{L} }{ 2 \Gamma _\text{L}  }   }
e^{- i a^{2}  \frac{\zeta_\text{L}  }{ 4}  }
e^{ -i \frac{ \epsilon_{} }{ {\cal E}_\text{L}  }  a \frac{\zeta_\text{L} }{ 2 } }
\nonumber \\
&&
\times
\left(  \int\limits _{0}^{ \infty } d \xi_\text{L}
e^{ - \xi_\text{L}  }
e^{-i \xi_\text{L} \frac{  \tau - \tau ^{\prime} }{ 2 \Gamma _\text{L}  }   }
e^{ -i a \xi_\text{L}  \frac{\zeta_\text{L}  }{ 2}  }  - 1 \right)
,
\label{a05}
\end{eqnarray}
and
\begin{eqnarray}
G_\text{L} = E_\text{L} + B_\text{L} &=&
\int\limits _{0}^{ \infty } d a
e^{ - \frac{ a }{2 } }
e^{-i a \frac{  \tau - \tau_\text{L}  }{ 2 \Gamma _\text{L}  }   }
e^{ i a^{2}  \frac{\zeta_\text{L}  }{ 4}  }
e^{ i \frac{ \epsilon_{} }{ {\cal E}_\text{L}  }  a \frac{\zeta_\text{L} }{ 2 } }
\nonumber \\
&&
\times
\left(  \int\limits _{0}^{ \infty } d \xi_\text{L} ^{\prime}
e^{ - \xi_\text{L} ^{\prime}  }
e^{-i \xi_\text{L}^{\prime}  \frac{ \tau - \tau^{\prime}  }{ 2 \Gamma _\text{L}  }   }
e^{ i a \xi_\text{L} ^{\prime}  \frac{\zeta_\text{L}  }{ 2}  }  - 1 \right)
.
\label{a06}
\end{eqnarray}

\subsection{The factor $P_\text{R}$}

Analogously, we transform the second factor and get
$P_\text{R} = F_\text{R}  + G_\text{R}$, with
\begin{eqnarray}
F_\text{R} &=&
\int\limits _{0}^{ \infty } d a
e^{ - \frac{ a }{2 } }
e^{i a  \frac{ t_{1} }{ 2 \Gamma _\text{R}  }   }
e^{ -i a^{2}  \frac{\zeta_\text{R}  }{ 4}  }
e^{ -i m \Omega 2 \Gamma _\text{R} a \frac{\zeta_\text{R} }{ 2 } }
e^{ -i \frac{ \epsilon_{} }{ {\cal E}_\text{R}  } a \frac{\zeta_\text{R} }{ 2 } }
\nonumber \\
&&
\times
\left(
\int\limits _{0}^{ \infty } d \xi_\text{R}
e^{ - \xi_\text{R}  }
e^{ - i \xi_\text{R} \left(  \frac{  t_{2} - t_{1} }{ 2 \Gamma _\text{R}  } +  a\frac{\zeta_\text{R} }{ 2 }  - \left(  n - m \right) \Omega 2 \Gamma _\text{R} \frac{\zeta_\text{R} }{ 2 }  \right)  }
- 1 \right)
,
\label{a07}
\end{eqnarray}
and
\begin{eqnarray}
G_\text{R} &=&
\int\limits _{0}^{ \infty } d a
e^{ - \frac{ a }{2 } }
e^{ - i a \frac{  t_{2} }{ 2 \Gamma _\text{R}  }   }
e^{ i a^{2}  \frac{\zeta_\text{R}  }{ 4}  }
e^{ i n \Omega 2 \Gamma _\text{R}  a \frac{\zeta_\text{R} }{ 2 } }
e^{ i \frac{ \epsilon_{} }{ {\cal E}_\text{R}  }  a \frac{\zeta_\text{R} }{ 2 } }
\nonumber \\
&&
\times
\left(
\int\limits _{0}^{ \infty } d \xi_\text{R} ^{\prime}
e^{ - \xi_\text{R} ^{\prime}  }
e^{- i \xi_\text{R}^{\prime} \left(   \frac{ t_{2} - t_{1} }{ 2 \Gamma _\text{R}  }   - a \frac{ \zeta_\text{R} }{ 2  } - (n-m) \Omega 2 \Gamma _\text{R}  \frac{\zeta }{ 2 } \right)  }
- 1 \right)
.
\label{a08}
\end{eqnarray}

\subsection{Collecting it all together}

With the above notation, Eq. (\ref{16}) of the main text becomes
\begin{eqnarray}
\delta G^{(1)}_\text{LR } &=&
\frac{ e^{  \frac{ i }{ \hbar  } \mu  \left(  t_{1} - t_{2} \right)  } }{ v_{ \mu} }
\sum\limits_{m,n}^{}
\iint\limits _{0}^{ {\cal T} _{0} }
\frac{ d \tau ^{\prime} }{ {\cal T} _{0}  }
\frac{ d \tau }{ {\cal T} _{0}  }
e^{-i m \Omega \left(  \tau ^{\prime} - t_{1} \right)}
e^{i n \Omega \left(  \tau - t_{2} \right)}
\nonumber \\
&& \times
\frac{ 1 }{ 4 \pi \Gamma _\text{R}^{}  }
\int\limits_{ 0 }^{ \infty }   d x
 e^{  i x \frac{  \left( t_{2} - t_{1} \right)  }{ 2 \Gamma _\text{R}^{}  } }
\left(  F_\text{L} + G_\text{L}  \right)
\left(  F_\text{R} + G_\text{R} \right)
,
\label{a09}
\end{eqnarray}
where we introduced $x = - \epsilon/ {\cal E}_\text{R}$.
Recall that for the zero temperature of interest here, the integral over energy $ \epsilon = E - \mu$ in Eq.~(\ref{16}) of the main text runs from $ - \infty$ up to $0$ with $f(E)=1$ throughout this interval.
Now we integrate over $x$.

\subsection{The energy integration}

\subsubsection{The term with the product $F_\text{L} F_\text{R}$}

In this case, we need to evaluate the next integral $I_\text{FF} = \int_{ 0 }^{ \infty }   d x  e^{  i x \frac{  \left( t_{2} - t_{1} \right)  }{ 2 \Gamma _\text{R}^{}  } }  F_\text{L} F_\text{R}$.
Recall that the limit $x \to \infty$ corresponds to $E \ll \mu$, where the exponential factor $e^{  i x \frac{  \left( t_{2} - t_{1} \right)  }{ 2 \Gamma _\text{R}^{}  } } $ should be treated as vanishing~\cite{Grenier:2011js,Ferraro:2013bt}.
So, we need to evaluate the following integral,
\begin{eqnarray}
I_\text{FF} &=& J_\text{FF}
\int\limits _{0}^{ \infty } d a_\text{L}
e^{ - \frac{ a_\text{L} }{2 } }
e^{i a_\text{L} \frac{ \tau ^{\prime} - \tau_\text{L} }{ 2 \Gamma _\text{L}  }   }
e^{- i a_\text{L}^{2}  \frac{\zeta_\text{L}  }{ 4}  }
\nonumber \\
&&
\times
\int\limits _{0}^{ \infty } d a_\text{R}
e^{ - \frac{ a_\text{R} }{2 } }
e^{  i a_\text{R} \frac{  t_{1} }{ 2 \Gamma _\text{R}  }   }
e^{ -i a_\text{R}^{2}  \frac{\zeta_\text{R}  }{ 4}  }
e^{ -i m \Omega 2 \Gamma _\text{R}  a_\text{R} \frac{\zeta_\text{R} }{ 2 } }
,
\label{a10}
\end{eqnarray}
with
\begin{eqnarray}
J_\text{FF} &=&
\int\limits_{ 0 }^{ \infty }   d x
 e^{  i x \left(  \frac{  \left( t_{2} - t_{1} \right)  }{ 2 \Gamma _\text{R}^{}  } + a_\text{L} C \frac{\zeta_\text{L} }{ 2 } + a_\text{R} \frac{\zeta_\text{R} }{ 2 }    \right)  }
\left(
\int\limits _{0}^{ \infty } d \xi_\text{L}
e^{ - \xi_\text{L}  }
e^{-i \xi_\text{L} \frac{  \tau - \tau ^{\prime} }{ 2 \Gamma _\text{L}  }   }
e^{ -i a_\text{L} \xi_\text{L}  \frac{\zeta_\text{L}  }{ 2}  }
- 1 \right)
\nonumber \\
&&
\times
\left(
\int\limits _{0}^{ \infty } d \xi_\text{R}
e^{ - \xi_\text{R}  }
e^{ - i \xi_\text{R} \left(  \frac{  t_{2} - t_{1} }{ 2 \Gamma _\text{R}  } +  a_\text{R}\frac{\zeta_\text{R} }{ 2 }  - \left(  n - m \right) \Omega 2 \Gamma _\text{R} \frac{\zeta_\text{R} }{ 2 }  \right)  }
- 1 \right)
.
\label{a11}
\end{eqnarray}
Here we introduced $C = \Gamma _\text{L}/ \Gamma _\text{R} = {\cal E}_\text{R}/ {\cal E}_\text{L}$.

After integrating out $x$ we get, $I_\text{FF} = I_\text{FF}^{(1)} + I_\text{FF}^{(2)}$, where
\begin{eqnarray}
I_\text{FF}^{(1)} &=&
\int\limits _{0}^{ \infty } d a_\text{L}
e^{ - \frac{ a_\text{L} }{2 } }
e^{i a_\text{L} \frac{ \tau ^{\prime} - \tau_\text{L} }{ 2 \Gamma _\text{L}  }   }
e^{- i a_\text{L}^{2}  \frac{\zeta_\text{L}  }{ 4}  }
\frac{ i\left(  \frac{ \tau - \tau ^{\prime} }{ 2 \Gamma _\text{L}  } + a_\text{L} \frac{ \zeta_\text{L} }{ 2  } \right) }{ -1 - i\left(  \frac{ \tau - \tau ^{\prime} }{ 2 \Gamma _\text{L}  } + a_\text{L} \frac{ \zeta_\text{L} }{ 2  } \right)  }
\nonumber \\
&&
\times
\int\limits _{0}^{ \infty } d a_\text{R}
e^{ - \frac{ a_\text{R} }{2 } }
e^{  i a_\text{R} \frac{  t_{1} }{ 2 \Gamma _\text{R}  }   }
e^{ -i a_\text{R}^{2}  \frac{\zeta_\text{R}  }{ 4}  }
e^{ -i m \Omega 2 \Gamma _\text{R}  a_\text{R} \frac{\zeta_\text{R} }{ 2 } }
\nonumber \\
&&
\times
\int\limits _{0}^{ \infty } d \xi_\text{R}
e^{ - \xi_\text{R}  }
e^{ - i \xi_\text{R} \left(  \frac{  t_{2} - t_{1} }{ 2 \Gamma _\text{R}  } +  a_\text{R}\frac{\zeta_\text{R} }{ 2 }  - \left(  n - m \right) \Omega 2 \Gamma _\text{R} \frac{\zeta_\text{R} }{ 2 }  \right)  }
,
\label{a12}
\end{eqnarray}
and
\begin{eqnarray}
I_\text{FF}^{(2)} &=&
\int\limits _{0}^{ \infty } d a_\text{L}
e^{ - \frac{ a_\text{L} }{2 } }
e^{i a_\text{L} \frac{ \tau ^{\prime} - \tau_\text{L} }{ 2 \Gamma _\text{L}  }   }
e^{- i a_\text{L}^{2}  \frac{\zeta_\text{L}  }{ 4}  }
\nonumber \\
&&
\times
\frac{ i\left(  \frac{ \tau - \tau ^{\prime} }{ 2 \Gamma _\text{L}  } + a_\text{L} \frac{ \zeta_\text{L} }{ 2  } \right) }{ -1 - i\left(  \frac{ \tau - \tau ^{\prime} }{ 2 \Gamma _\text{L}  } + a_\text{L} \frac{ \zeta_\text{L} }{ 2  } \right)  }
\frac{
- i\left(  n - m \right) \Omega 2 \Gamma _\text{R} \frac{\zeta_\text{R} }{ 2 }
- i a_\text{L} C \frac{ \zeta_\text{L} }{ 2  }
}{  i \left(  \frac{  \left( t_{2} - t_{1} \right)  }{ 2 \Gamma _\text{R}^{}  } + a_\text{L} C  \frac{\zeta_\text{L} }{ 2 } + a_\text{R} \frac{\zeta_\text{R} }{ 2 }    \right)  }
\nonumber \\
&&
\times
\int\limits _{0}^{ \infty } d a_\text{R}
e^{ - \frac{ a_\text{R} }{2 } }
e^{  i a_\text{R} \frac{  t_{1} }{ 2 \Gamma _\text{R}  }   }
e^{ -i a_\text{R}^{2}  \frac{\zeta_\text{R}  }{ 4}  }
e^{ -i m \Omega 2 \Gamma _\text{R}  a_\text{R} \frac{\zeta_\text{R} }{ 2 } }
\nonumber \\
&&
\times
\int\limits _{0}^{ \infty } d \xi_\text{R}
e^{ - \xi_\text{R}  }
e^{ - i \xi_\text{R} \left(  \frac{  t_{2} - t_{1} }{ 2 \Gamma _\text{R}  } +  a_\text{R}\frac{\zeta_\text{R} }{ 2 }  - \left(  n - m \right) \Omega 2 \Gamma _\text{R} \frac{\zeta_\text{R} }{ 2 }  \right)  } .
\label{a13}
\end{eqnarray}
To simplify $I_\text{FF}^{(2)}$, let us remember that this term is under  summation over $n$ and $m$ in Eq.~(\ref{a09}).
This summation together with integrations over $ \tau$ and $ \tau ^{\prime}$ gives us $\tau - \tau ^{\prime} = t_{2} - t_{1} + a_\text{R} \Gamma _\text{R} \zeta_\text{R}$.
Using this relation, we find
\begin{eqnarray}
I_\text{FF}^{(2)} &=&
\frac{ 1 }{ C }
\int\limits _{0}^{ \infty } d a_\text{L}
e^{ - \frac{ a_\text{L} }{2 } }
e^{i a_\text{L} \frac{ \tau ^{\prime} - \tau_\text{L} }{ 2 \Gamma _\text{L}  }   }
e^{- i a_\text{L}^{2}  \frac{\zeta_\text{L}  }{ 4}  }
\frac{
- i\left(  n - m \right) \Omega 2 \Gamma _\text{R} \frac{\zeta_\text{R} }{ 2 }
- i a_\text{L} C \frac{ \zeta_\text{L} }{ 2  }
}{  -1 - i\left(  \frac{ \tau - \tau ^{\prime} }{ 2 \Gamma _\text{L}  } + a_\text{L} \frac{ \zeta_\text{L} }{ 2  } \right)  }
\nonumber \\
&&
\times
\int\limits _{0}^{ \infty } d a_\text{R}
e^{ - \frac{ a_\text{R} }{2 } }
e^{  i a_\text{R} \frac{  t_{1} }{ 2 \Gamma _\text{R}  }   }
e^{ -i a_\text{R}^{2}  \frac{\zeta_\text{R}  }{ 4}  }
e^{ -i m \Omega 2 \Gamma _\text{R}  a_\text{R} \frac{\zeta_\text{R} }{ 2 } }
\nonumber \\
&&
\times
\int\limits _{0}^{ \infty } d \xi_\text{R}
e^{ - \xi_\text{R}  }
e^{ - i \xi_\text{R} \left(  \frac{  t_{2} - t_{1} }{ 2 \Gamma _\text{R}  } +  a_\text{R}\frac{\zeta_\text{R} }{ 2 }  - \left(  n - m \right) \Omega 2 \Gamma _\text{R} \frac{\zeta_\text{R} }{ 2 }  \right)  } .
\label{a14}
\end{eqnarray}
Now, the sum $I_\text{FF}^{} = I_\text{FF}^{(1)} + I_\text{FF}^{(2)}$, becomes
\begin{eqnarray}
I_\text{FF}^{} &=&
\frac{ 1 }{ C }
\left(
i\left(  n - m \right) \Omega 2 \Gamma _\text{R} \frac{\zeta_\text{R} }{ 2 }
-i \frac{ \tau - \tau ^{\prime} }{ 2 \Gamma _\text{R}  }
\right)
\nonumber \\
&&
\times
\int\limits _{0}^{ \infty } d a_\text{L}
e^{ - \frac{ a_\text{L} }{2 } }
e^{i a_\text{L} \frac{ \tau ^{\prime} - \tau_\text{L} }{ 2 \Gamma _\text{L}  }   }
e^{- i a_\text{L}^{2}  \frac{\zeta_\text{L}  }{ 4}  }
\nonumber \\
&&
\times
\int\limits _{0}^{ \infty } d \xi_\text{L}
e^{ - \xi_\text{L}  }
e^{-i \xi_\text{L} \frac{  \tau - \tau ^{\prime} }{ 2 \Gamma _\text{L}  }   }
e^{ -i a_\text{L} \xi_\text{L}  \frac{\zeta_\text{L}  }{ 2}  }
\nonumber \\
&&
\times
\int\limits _{0}^{ \infty } d a_\text{R}
e^{ - \frac{ a_\text{R} }{2 } }
e^{  i a_\text{R} \frac{  t_{1} }{ 2 \Gamma _\text{R}  }   }
e^{ -i a_\text{R}^{2}  \frac{\zeta_\text{R}  }{ 4}  }
e^{ -i m \Omega 2 \Gamma _\text{R}  a_\text{R} \frac{\zeta_\text{R} }{ 2 } }
\nonumber \\
&&
\times
\int\limits _{0}^{ \infty } d \xi_\text{R}
e^{ - \xi_\text{R}  }
e^{ - i \xi_\text{R} \left(  \frac{  t_{2} - t_{1} }{ 2 \Gamma _\text{R}  } +  a_\text{R}\frac{\zeta_\text{R} }{ 2 }  - \left(  n - m \right) \Omega 2 \Gamma _\text{R} \frac{\zeta_\text{R} }{ 2 }  \right)  }
,
\label{a15}
\end{eqnarray}
where we represented the denominator as an integral over $ \xi_\text{L}$.

Finally, we go over from $a_{j} = \xi_{j} ^{\prime} - \xi_{j}$, $j = \text{L, R}$, back to $ \xi ^{\prime}_{j}$,
\begin{eqnarray}
I_\text{FF}^{} &=&
\frac{ 1 }{ C }
\left(
i\left(  n - m \right) \Omega 2 \Gamma _\text{R} \frac{\zeta_\text{R} }{ 2 }
-i \frac{ \tau - \tau ^{\prime} }{ 2 \Gamma _\text{R}  }
\right)
\nonumber \\
&&
\times
\int\limits _{ \xi_\text{L}  }^{ \infty } d \xi_\text{L} ^{\prime}
e^{ - \frac{ \xi_\text{L} ^{\prime} }{2 } }
e^{i \xi_\text{L}^{\prime}  \frac{ \tau ^{\prime} - \tau_\text{L} }{ 2 \Gamma _\text{L}  }   }
e^{- i \left(  \xi_\text{L}^{\prime} \right)^{2}  \frac{\zeta_\text{L}  }{ 4}  }
\nonumber \\
&&
\times
\int\limits _{0}^{ \infty } d \xi_\text{L}
e^{ - \frac{ \xi_\text{L} }{2 } }
e^{-i \xi_\text{L} \frac{  \tau - \tau_\text{L}  }{ 2 \Gamma _\text{L}  }   }
e^{ i \xi_\text{L}^{2}  \frac{\zeta_\text{L}  }{ 4}  }
\nonumber \\
&&
\times
\int\limits _{ \xi_\text{R} }^{ \infty } d \xi_\text{R} ^{\prime}
e^{ - \frac{ \xi_\text{R} ^{\prime} }{2 } }
e^{i \xi_\text{R}^{\prime}  \frac{ t_{1} }{ 2 \Gamma _\text{R}  }   }
e^{- i \left(  \xi_\text{R}^{\prime} \right)^{2}  \frac{\zeta_\text{R}  }{ 4}  }
e^{ -i m \Omega 2 \Gamma _\text{R} \xi_\text{R}^{\prime} \frac{\zeta_\text{R} }{ 2 } }
\nonumber \\
&&
\times
\int\limits _{0}^{ \infty } d \xi_\text{R}
e^{ - \frac{ \xi_\text{R} }{2 } }
e^{-i \xi_\text{R} \frac{  t_{2} }{ 2 \Gamma _\text{R}  }   }
e^{ i \xi_\text{R}^{2}  \frac{\zeta_\text{R}  }{ 4}  }
e^{ i n \Omega 2 \Gamma _\text{R} \xi_\text{R} \frac{\zeta_\text{R} }{ 2 } }
.
\label{a16}
\end{eqnarray}

\subsubsection{The other terms}

The terms with factors $F_\text{L}G_\text{R}$, $G_\text{L}F_\text{R}$, and $G_\text{L}G_\text{R}$ in Eq.~(\ref{a09}) can be represented as Eq.~(\ref{a16}), but with different limits of integration.
Together, they cover the area of integration over various $ \xi$'s from $0$ to $ \infty$.
Then, the additional contribution $ \delta G ^{(1)}_\text{LR}$, Eq.~(\ref{a09}), reads
\begin{eqnarray}
\delta G^{(1)}_\text{LR } &=&
\frac{ 1 }{ C }
\sum\limits_{m,n}^{}
\iint\limits _{0}^{ {\cal T} _{0} }
\frac{ d \tau ^{\prime} d \tau }{ {\cal T} _{0}^{2}  }
\frac{
e^{-i m \Omega \left(  \tau ^{\prime} - t_{1} \right)}
e^{i n \Omega \left(  \tau - t_{2} \right)}
}{ 4 \pi \Gamma _\text{R}^{} v_{ \mu} }
\left(
i\left(  n - m \right) \Omega 2 \Gamma _\text{R} \frac{\zeta_\text{R} }{ 2 }
-i \frac{ \tau - \tau ^{\prime} }{ 2 \Gamma _\text{R}  }
\right)
\nonumber \\
&&
\times
\int\limits _{0  }^{ \infty } d \xi_\text{L} ^{\prime}
e^{ - \frac{ \xi_\text{L} ^{\prime} }{2 } }
e^{i \xi_\text{L}^{\prime}  \frac{ \tau ^{\prime} - \tau_\text{L} }{ 2 \Gamma _\text{L}  }   }
e^{- i \left(  \xi_\text{L}^{\prime} \right)^{2}  \frac{\zeta_\text{L}  }{ 4}  }
\nonumber \\
&&
\times
\int\limits _{0}^{ \infty } d \xi_\text{L}
e^{ - \frac{ \xi_\text{L} }{2 } }
e^{-i \xi_\text{L} \frac{  \tau - \tau_\text{L}  }{ 2 \Gamma _\text{L}  }   }
e^{ i \xi_\text{L}^{2}  \frac{\zeta_\text{L}  }{ 4}  }
\nonumber \\
&&
\times
\int\limits _{ 0 }^{ \infty } d \xi_\text{R} ^{\prime}
e^{ - \frac{ \xi_\text{R} ^{\prime} }{2 } }
e^{i \xi_\text{R}^{\prime}  \frac{ t_{1} }{ 2 \Gamma _\text{R}  }   }
e^{- i \left(  \xi_\text{R}^{\prime} \right)^{2}  \frac{\zeta_\text{R}  }{ 4}  }
e^{ -i m \Omega 2 \Gamma _\text{R} \xi_\text{R}^{\prime} \frac{\zeta_\text{R} }{ 2 } }
\nonumber \\
&&
\times
\int\limits _{0}^{ \infty } d \xi_\text{R}
e^{ - \frac{ \xi_\text{R} }{2 } }
e^{-i \xi_\text{R} \frac{  t_{2} }{ 2 \Gamma _\text{R}  }   }
e^{ i \xi_\text{R}^{2}  \frac{\zeta_\text{R}  }{ 4}  }
e^{ i n \Omega 2 \Gamma _\text{R} \xi_\text{R} \frac{\zeta_\text{R} }{ 2 } }
.
\label{a17}
\end{eqnarray}

\subsection{The inverse Fourier transformation}

To simplify the terms with $n$ and $m$, we use  $ - i m \Omega e^{- i m \Omega \tau ^{\prime} } \to \frac{  \partial  }{  \partial \tau ^{\prime}  } e^{- i m \Omega \tau ^{\prime} }$ and $  i n \Omega e^{ i n \Omega \tau  } \to \frac{  \partial  }{  \partial \tau  } e^{ i n \Omega \tau  }$.
The integration by parts over $ \tau$ and $ \tau ^{\prime}$ results in the following
\begin{eqnarray}
\left[
i\left(  n - m \right) \Omega 2 \Gamma _\text{R} \frac{\zeta_\text{R} }{ 2 }
-i \frac{ \tau - \tau ^{\prime} }{ 2 \Gamma _\text{R}  }
\right]
\!\Rightarrow
-
\frac{ 1 }{ C }
\left(
\left[ \frac{  \partial  }{  \partial \tau ^{\prime}  } + \frac{  \partial  }{  \partial \tau  } \right] 2 \Gamma _\text{R} \frac{\zeta_\text{R} }{ 2 }
+i \frac{ \tau - \tau ^{\prime} }{ 2 \Gamma _\text{R}  }
\right) .
\label{a19}
\end{eqnarray}
After taking the derivative over $ \tau$ and $ \tau ^{\prime}$, we perform  summation over $n$ and $m$ and integration over $ \tau$ and $ \tau ^{\prime}$ in Eq.~(\ref{a17}), using $\tau ^{\prime} = t_{1} - \xi_\text{R} ^{\prime} \Gamma _\text{R} \zeta_\text{R}$ and $ \tau = t_{2} - \xi_\text{R}  \Gamma _\text{R} \zeta_\text{R}$, and get
\begin{eqnarray}
\delta G^{(1)}_\text{LR } &=&
\int\limits _{ 0  }^{ \infty } d \xi_\text{L} ^{\prime}
e^{ - \frac{ \xi_\text{L} ^{\prime} }{2 } }
e^{i \xi_\text{L}^{\prime}  \frac{ t_{1} - \tau_\text{L} }{ 2 \Gamma _\text{L}  }   }
e^{- i \left(  \xi_\text{L}^{\prime} \right)^{2}  \frac{\zeta_\text{L}  }{ 4}  }
e^{-i \frac{ \xi_\text{L}^{\prime} \xi_\text{R} ^{\prime} }{ C }  \frac{ \zeta_\text{R} }{ 2  }   }
\nonumber \\
&&
\times
\int\limits _{0}^{ \infty } d \xi_\text{L}
e^{ - \frac{ \xi_\text{L} }{2 } }
e^{-i \xi_\text{L} \frac{  t_{2} - \tau_\text{L}  }{ 2 \Gamma _\text{L}  }   }
e^{ i \xi_\text{L}^{2}  \frac{\zeta_\text{L}  }{ 4}  }
e^{i \frac{ \xi_\text{L} \xi_\text{R} }{ C } \frac{  \zeta_\text{R}  }{ 2   }   }
\nonumber \\
&&
\times
\frac{
1
}{ 4 \pi \Gamma _\text{L}^{} v_{ \mu} }
\int\limits _{ 0 }^{ \infty } d \xi_\text{R} ^{\prime}
e^{ - \frac{ \xi_\text{R} ^{\prime} }{2 } }
e^{i \xi_\text{R}^{\prime}  \frac{ t_{1} }{ 2 \Gamma _\text{R}  }   }
e^{- i \left(  \xi_\text{R}^{\prime} \right)^{2}  \frac{\zeta_\text{R}  }{ 4}  }
\nonumber \\
&&
\times
\int\limits _{0}^{ \infty } d \xi_\text{R}
e^{ - \frac{ \xi_\text{R} }{2 } }
e^{-i \xi_\text{R} \frac{  t_{2} }{ 2 \Gamma _\text{R}  }   }
e^{ i \xi_\text{R}^{2}  \frac{\zeta_\text{R}  }{ 4}  }
\nonumber \\
&&
\times
\left(
i \frac{ t_{1} - t_{2} }{ 2 \Gamma _\text{R}  }
+ i \Gamma _\text{R} \frac{ \zeta_\text{R} }{ 2 }
\left[
\frac{ \left(  \xi_\text{L} - \xi_\text{L} ^{\prime} \right) }{ \Gamma _\text{L} }
+
\frac{ \left(  \xi_\text{R} - \xi_\text{R} ^{\prime} \right) }{ \Gamma _\text{R}  }
\right]
\right)
.
\label{a20}
\end{eqnarray}

\subsection{Further processing}

The last factor in Eq.~(\ref{a20}) can be formally represented as follows,
\begin{eqnarray}
i \frac{ t_{1} - t_{2} }{ 2 \Gamma _\text{R}  }
+ i  \frac{ \zeta_\text{R} }{ 2 }
\left[
\frac{ \left(  \xi_\text{L} - \xi_\text{L} ^{\prime} \right) }{ C }
+
\left(  \xi_\text{R} - \xi_\text{R} ^{\prime} \right)
\right]
\Rightarrow
1 + \frac{  \partial }{  \partial \xi_\text{R} ^{\prime}  } + \frac{  \partial }{  \partial \xi_\text{R}  } ,
\label{a21}
\end{eqnarray}
where differentiation is applied on the whole integrand.
As a result, we have three contributions, $\delta G^{(1)}_\text{LR} = \delta G^{(1)}_\text{LR,1} + \delta G^{(1)}_\text{LR,2} + \delta G^{(1)}_\text{LR,3}$.

To evaluate the term with $  \partial/  \partial \xi_\text{R} ^{\prime}$, we integrate over $ \xi_\text{R} ^{\prime}$ by parts,  resulting in $-1$, and get
\begin{eqnarray}
\delta G^{(1)}_\text{LR,1 } &=&
\frac{
-1
}{ 4 \pi \Gamma _\text{L}^{} v_{ \mu} }
\int\limits _{ 0  }^{ \infty } d \xi_\text{L} ^{\prime}
e^{ - \frac{ \xi_\text{L} ^{\prime} }{2 } }
e^{i \xi_\text{L}^{\prime}  \frac{ t_{1} - \tau_\text{L} }{ 2 \Gamma _\text{L}  }   }
e^{- i \left(  \xi_\text{L}^{\prime} \right)^{2}  \frac{\zeta_\text{L}  }{ 4}  }
\nonumber \\
&&
\times
\int\limits _{0}^{ \infty } d \xi_\text{L}
e^{ - \frac{ \xi_\text{L} }{2 } }
e^{-i \xi_\text{L} \frac{  t_{2} - \tau_\text{L}  }{ 2 \Gamma _\text{L}  }   }
e^{ i \xi_\text{L}^{2}  \frac{\zeta_\text{L}  }{ 4}  }
e^{i \frac{ \xi_\text{L} \xi_\text{R} }{ C } \frac{  \zeta_\text{R}  }{ 2   }   }
\nonumber \\
&&
\times
\int\limits _{0}^{ \infty } d \xi_\text{R}
e^{ - \frac{ \xi_\text{R} }{2 } }
e^{-i \xi_\text{R} \frac{  t_{2} }{ 2 \Gamma _{\tau}  }   }
e^{ i \xi_\text{R}^{2}  \frac{\zeta  }{ 4}  }
.
\label{a22}
\end{eqnarray}
Similarly, we evaluate the term with $  \partial/  \partial \xi_\text{R}$,
\begin{eqnarray}
\delta G^{(1)}_\text{LR,2 } &=&
\int\limits _{ 0  }^{ \infty } d \xi_\text{L} ^{\prime}
e^{ - \frac{ \xi_\text{L} ^{\prime} }{2 } }
e^{i \xi_\text{L}^{\prime}  \frac{ t_{1} - \tau_\text{L} }{ 2 \Gamma _\text{L}  }   }
e^{- i \left(  \xi_\text{L}^{\prime} \right)^{2}  \frac{\zeta_\text{L}  }{ 4}  }
e^{-i \frac{ \xi_\text{L}^{\prime} \xi_\text{R}  ^{\prime} }{ C } \frac{ \zeta_\text{R} }{ 2  }   }
\nonumber \\
&&
\times
\int\limits _{0}^{ \infty } d \xi_\text{L}
e^{ - \frac{ \xi_\text{L} }{2 } }
e^{-i \xi_\text{L} \frac{  t_{2} - \tau_\text{L}  }{ 2 \Gamma _\text{L}  }   }
e^{ i \xi_\text{L}^{2}  \frac{\zeta_\text{L}  }{ 4}  }
\nonumber \\
\nonumber \\
&&
\times
\frac{
\left( -1 \right)
}{ 4 \pi \Gamma _\text{L}^{} v_{ \mu} }
\int\limits _{ 0 }^{ \infty } d \xi_\text{R} ^{\prime}
e^{ - \frac{ \xi_\text{R} ^{\prime} }{2 } }
e^{i \xi_\text{R}^{\prime}  \frac{ t_{1} }{ 2 \Gamma _\text{R}  }   }
e^{- i \left(  \xi_\text{R}^{\prime} \right)^{2}  \frac{\zeta_\text{R}  }{ 4}  }
.
\label{a23}
\end{eqnarray}
And finally, the term with $1$ in Eq.~(\ref{a21}) results in the following
\begin{eqnarray}
\delta G^{(1)}_\text{LR,3 } &=&
\int\limits _{ 0  }^{ \infty } d \xi_\text{L} ^{\prime}
e^{ - \frac{ \xi_\text{L} ^{\prime} }{2 } }
e^{i \xi_\text{L}^{\prime}  \frac{ t_{1} - \tau_\text{L} }{ 2 \Gamma _\text{L}  }   }
e^{- i \left(  \xi_\text{L}^{\prime} \right)^{2}  \frac{\zeta_\text{L}  }{ 4}  }
e^{-i \frac{ \xi_\text{L}^{\prime} \xi_\text{R} ^{\prime} }{ C }  \frac{ \zeta_\text{R} }{ 2  }   }
\nonumber \\
&&
\times
\int\limits _{0}^{ \infty } d \xi_\text{L}
e^{ - \frac{ \xi_\text{L} }{2 } }
e^{-i \xi_\text{L} \frac{  t_{2} - \tau_\text{L}  }{ 2 \Gamma _\text{L}  }   }
e^{ i \xi_\text{L}^{2}  \frac{\zeta_\text{L}  }{ 4}  }
e^{i \frac{ \xi_\text{L} \xi_\text{R} }{ C } \frac{  \zeta_\text{R}  }{ 2   }   }
\nonumber \\
&&
\times
\frac{
1
}{ 4 \pi \Gamma _\text{L}^{} v_{ \mu} }
\int\limits _{ 0 }^{ \infty } d \xi_\text{R} ^{\prime}
e^{ - \frac{ \xi_\text{R} ^{\prime} }{2 } }
e^{i \xi_\text{R}^{\prime}  \frac{ t_{1} }{ 2 \Gamma _\text{R}  }   }
e^{- i \left(  \xi_\text{R}^{\prime} \right)^{2}  \frac{\zeta_\text{R}  }{ 4}  }
\nonumber \\
&&
\times
\int\limits _{0}^{ \infty } d \xi_\text{R}
e^{ - \frac{ \xi_\text{R} }{2 } }
e^{-i \xi_\text{R} \frac{  t_{2} }{ 2 \Gamma _\text{R}  }   }
e^{ i \xi_\text{R}^{2}  \frac{\zeta_\text{R}  }{ 4}  }
.
\label{a24}
\end{eqnarray}
Combining $ \delta G ^{(1)}_\text{LR}$, given by the sum of Eqs.~(\ref{a22}) - (\ref{a24}), with $G ^{(1)}_\text{L}$, Eq.~(\ref{a02}) for $j=\text{L}$, we represent  the scattering amplitude of the composite source as follows, $G ^{(1)}_\text{tot} = G ^{(1)}_\text{R} + G ^{(1)}_\text{LR}$, where
\begin{eqnarray}
G ^{(1)}_\text{LR}\left( t_{1};t_{2} \right) =
 \frac{ e^{  \frac{ i }{ \hbar  } \mu  \left(  t_{1} - t_{2} \right)  } }{ v_{ \mu}  }
\psi_{2}^{*}\left(  t_{1} \right)
\psi_{2}^{}\left(  t_{2} \right).
\label{a25}
\end{eqnarray}
The function $ \psi_{2}\left(  t \right)$ is given in Eq.~(\ref{27}) of the main text with $ \eta_\text{L} = \eta_\text{R} = +1$ and $ \tau_\text{R} =0$.

\section{Normalization of the wave function $ \psi_{2}$, Eq.~(\ref{27}) of the main text}
\label{appB}

Here we show that the following function is normalized,
\begin{eqnarray}
\psi_{2}\left(  t \right) &=&
\frac{  1 }{  \sqrt{\pi  \Gamma _\text{L} } }
\int\limits_{0}^{ \infty} d \xi
e^{ -   \xi  }
e^{  - i \xi \frac{ t - t_\text{L} }{ \Gamma _\text{L}  }    }
e^{ i \xi^{2} \zeta_\text{L}  }
\label{b00} \\
&&
-\frac{  1 }{  2 \sqrt{\pi  \Gamma _\text{L} } }
\int\limits _{0}^{ \infty } d \xi_\text{L}
e^{ - \frac{ \xi_\text{L} }{2 } }
e^{-i \xi_\text{L} \frac{  t_{} - t_\text{L}  }{ 2 \Gamma _\text{L}  }   }
e^{ i \xi_\text{L}^{2}  \frac{\zeta_\text{L}  }{ 4}  }
\int\limits _{0}^{ \infty } d \xi_\text{R}
e^{ - \frac{ \xi_\text{R} }{2 } }
e^{-i \xi_\text{R} \frac{  t_{} }{ 2 \Gamma _\text{R}  }   }
e^{ i \xi_\text{R}^{2}  \frac{\zeta_\text{R}  }{ 4}  }
e^{i  \xi_\text{L} \xi_\text{R}  \frac{  \zeta_\text{R}  }{ 2 C   }   }
.
\nonumber
\end{eqnarray}
Notice some change of integration variables compared to original Eq.~(\ref{27}) of the main text.
For definiteness, we consider the case of $ \eta_\text{L} = \eta_\text{R} = +1$.
The other cases are analysed in a similar way.
Recall that here we set $ \tau_\text{R} =0$.

We need to evaluate the following integral ($x = t/ \Gamma _\text{L}$)
\begin{eqnarray}
\int\limits _{ - \infty}^{ \infty } dx \left | \psi_{2}\left( \Gamma _{\tau} x \right) \right |^{2} = I_{1} + I_{2} + I_{3},
\label{b01}
\end{eqnarray}
which we represented as the sum of three terms.
The first term is the following $\left(  x_\text{L} = \tau_\text{L} / \Gamma _\text{L} \right)$,
\begin{eqnarray}
I_{1} =
\frac{ 1 }{ \pi  }
\int\limits _{ - \infty}^{ \infty } dx
\left |
\int\limits_{0}^{ \infty} d \xi
e^{ -   \xi_{}  }
e^{  - i \left(  t - x_\text{L}  \right) \xi }
e^{ i  \zeta_\text{L} \xi^{2} }
 \right |^{2} = 1.
\label{b02}
\end{eqnarray}
This is evaluated trivially and represents the fact that the wave function injected by a single source, $ \psi_{1}$, Eq.~(\ref{24}) of the main text, is normalized.
To demonstrate that $ \psi_{2}$, Eq.~(\ref{b00}), is also normalized, we need to show that $I_{2} = - I_{3}$.

\subsection{The term $I_{2}$}
First, we evaluate
\begin{eqnarray}
I_{2} &=&
-  \frac{ 1 }{ \pi  } {\rm Re}
\int\limits _{ - \infty}^{ \infty } dx
\int\limits_{0}^{ \infty} d \xi_{}
e^{ -   \xi_{}  }
e^{   i \xi_{}  \left(  x - x_\text{L} \right)      }
e^{ -i \xi_{}^{2} \zeta_\text{L}  }
\nonumber \\
&&
\times
\int\limits _{0}^{ \infty } d \xi_\text{L}
e^{ - \frac{ \xi_\text{L} }{2 } }
e^{-i \xi_\text{L} \frac{  x_{} - x_\text{L}  }{ 2   }   }
e^{ i \xi_\text{L}^{2}  \frac{\zeta_\text{L}  }{ 4}  }
e^{i  \xi_\text{L} \xi_\text{R}  \frac{  \zeta_\text{R}  }{ 2 C   }   }
\int\limits _{0}^{ \infty } d \xi_\text{R}
e^{ - \frac{ \xi_\text{R} }{2 } }
e^{-i \xi_\text{R} x \frac{ C }{ 2  }   }
e^{ i \xi_\text{R}^{2}  \frac{\zeta_\text{R}  }{ 4}  }
.
\label{b03}
\end{eqnarray}
The integral over $x$ gives us
\begin{eqnarray}
I_{2} &=&
-  \frac{ 1 }{ \pi  } {\rm Re}
\int\limits_{0}^{ \infty} d \xi_{}
e^{ -   \xi_{}  }
e^{  - i \xi_{}  x_\text{L}      }
e^{ -i \xi_{}^{2} \zeta_\text{L}  }
\int\limits _{0}^{ \infty } d \xi_\text{L}
e^{ - \frac{ \xi_\text{L} }{2 } }
e^{  i \xi_\text{L} \frac{  x_\text{L}  }{ 2   }   }
e^{ i \xi_\text{L}^{2}  \frac{\zeta_\text{L}  }{ 4}  }
e^{i  \xi_\text{L} \xi_\text{R}  \frac{  \zeta_\text{R}  }{ 2 C   }   }
\nonumber \\
&&
\times
\int\limits _{0}^{ \infty } d \xi_\text{R}
e^{ - \frac{ \xi_\text{R} }{2 } }
e^{ i \xi_\text{R}^{2}  \frac{\zeta_\text{R}  }{ 4}  }
2 \pi \delta\left( \xi_{} - \xi_\text{R} \frac{ C }{ 2 } - \xi_\text{L}\frac{ 1 }{ 2 } \right)
.
\label{b04}
\end{eqnarray}
Recall that $C = \Gamma _\text{L}/ \Gamma _\text{R}$.
To get rid of the $\delta$-function, we integrate over $ \xi_{}$ and use  $ \xi_{} = \xi_\text{R} \frac{ C }{ 2 } + \xi_\text{L} \frac{ 1 }{ 2 }$.
Note that such integration can always be done, because $ \xi_{} > 0$ for any $ \xi_\text{R}>0$, $ \xi_\text{L}>0$, and $C>0$.
As a result, we have,
\begin{eqnarray}
I_{2} &=&
-  2 {\rm Re}
\int\limits _{0}^{ \infty } d \xi_\text{R}
e^{ - \xi_\text{R}\frac{ 1+ C  }{2 } }
e^{ i \xi_\text{R}^{2}  \frac{\zeta_\text{R} - C^{2} \zeta_\text{L}  }{ 4}  }
e^{  - i \xi_\text{R} \frac{ C x_\text{L} }{ 2 }      }
\int\limits _{0}^{ \infty } d \xi_\text{L}
e^{ - \xi_\text{L}  }
e^{i  \xi_\text{L} \xi_\text{R}  \frac{  \zeta_\text{R} - C^{2} \zeta_\text{L}  }{ 2 C   }   }
.
\label{b05}
\end{eqnarray}
We can now carry out the integral over $ \xi_\text{L}$ and get
\begin{eqnarray}
I_{2} &=&
-  2 {\rm Re}
\int\limits _{0}^{ \infty } d \xi_\text{R}
\frac{
e^{ - \xi_\text{R}\frac{ 1+ C  }{2 } }
e^{ i \xi_\text{R}^{2}  \frac{\zeta_\text{R} - C^{2} \zeta_\text{L}  }{ 4}  }
e^{  - i \xi_\text{R} \frac{ C x_\text{L} }{ 2 }      }
}{ 1  - i \xi_\text{R} \frac{ \zeta_\text{R} - C^{2} \zeta_\text{L} }{ 2 C }  }
.
\label{b06}
\end{eqnarray}

\subsection{The term $I_{3}$}

\begin{eqnarray}
I_{3}
&=&
\frac{ 1 }{ 4 \pi  }
\int\limits _{ - \infty}^{ \infty } dx
\int\limits _{0}^{ \infty } d \xi_\text{R}
e^{ - \frac{ \xi_\text{R} }{2 } }
e^{-i x \xi_\text{R} \frac{  C }{ 2   }   }
e^{ i \xi_\text{R}^{2}  \frac{\zeta_\text{R}  }{ 4}  }
\nonumber \\
&&
\times
\int\limits _{0}^{ \infty } d \xi_\text{L}
e^{ - \frac{ \xi_\text{L} }{2 } }
e^{-i \xi_\text{L} \frac{  x - x_\text{L}  }{ 2   }   }
e^{ i \xi_\text{L}^{2}  \frac{\zeta_\text{L}  }{ 4}  }
e^{i  \xi_\text{L} \xi_\text{R}  \frac{  \zeta_\text{R}  }{ 2 C   }   }
\nonumber \\
&&
\times
\int\limits _{0}^{ \infty } d \chi_\text{R}
e^{ - \frac{ \chi_\text{R} }{2 } }
e^{i x \chi_\text{R} \frac{  C}{ 2   }   }
e^{- i \chi_\text{R}^{2}  \frac{\zeta_\text{R}  }{ 4}  }
\nonumber \\
&&
\times
\int\limits _{0}^{ \infty } d \chi_\text{L}
e^{ - \frac{ \chi_\text{L} }{2 } }
e^{i \chi_\text{L} \frac{  x - x_\text{L}  }{ 2   }   }
e^{- i \chi_\text{L}^{2}  \frac{\zeta_\text{L}  }{ 4}  }
e^{-i  \chi_\text{L} \chi_\text{R}  \frac{  \zeta_\text{R}  }{ 2 C   }   } .
\label{b07}
\end{eqnarray}
After integrating out $x$ we have,
\begin{eqnarray}
I_{3} &=&
\frac{ 1 }{ 4 \pi  }
\int\limits _{0}^{ \infty } d \xi_\text{R}
e^{ - \frac{ \xi_\text{R} }{2 } }
e^{ i \xi_\text{R}^{2}  \frac{\zeta_\text{R}  }{ 4}  }
\int\limits _{0}^{ \infty } d \chi_\text{R}
e^{ - \frac{ \chi_\text{R} }{2 } }
e^{- i \chi_\text{R}^{2}  \frac{\zeta_\text{R}  }{ 4}  }
\nonumber \\
&&
\times
\int\limits _{0}^{ \infty } d \xi_\text{L}
e^{ - \frac{ \xi_\text{L} }{2 } }
e^{i \xi_\text{L} \frac{  x_\text{L}  }{ 2   }   }
e^{ i \xi_\text{L}^{2}  \frac{\zeta_\text{L}  }{ 4}  }
e^{i  \xi_\text{L} \xi_\text{R}  \frac{  \zeta_\text{R}  }{ 2 C   }   }
\nonumber \\
&&
\times
\int\limits _{0}^{ \infty } d \chi_\text{L}
e^{ - \frac{ \chi_\text{L} }{2 } }
e^{-i \chi_\text{L} \frac{   x_\text{L}  }{ 2   }   }
e^{- i \chi_\text{L}^{2}  \frac{\zeta_\text{L}  }{ 4}  }
e^{-i  \chi_\text{L} \chi_\text{R}  \frac{  \zeta_\text{R}  }{ 2 C   }   }
\nonumber \\
&&
\times
2 \pi \delta\left(  \xi_\text{R} \frac{ C }{ 2 } + \xi_\text{L}\frac{ 1 }{ 2 } - \chi_\text{R}\frac{ C }{ 2 } - \chi_\text{L}\frac{ 1 }{ 2 } \right)
.
\label{b08}
\end{eqnarray}
To get rid of the $\delta$-function, we integrate over $ \xi_\text{L}$ and use
$ \xi_\text{L} = - C \xi_\text{R} + C \chi_\text{R} + \chi_\text{L}$.
The condition $ \xi_\text{L} > 0$ demands $ \xi_\text{R} < \chi_\text{R} + \chi_\text{L}/C$.
For the sake of simplicity, we rescale $ \chi_\text{L} \to C \chi_\text{L}$, and shift $ \xi_\text{R} \to \xi_\text{R} + \chi_\text{R}$.
As a result $I_{3}$ becomes
\begin{eqnarray}
I_{3} &=& C
\int\limits _{0}^{ \infty } d \chi_\text{R}
e^{ - \chi_\text{R}  }
\int\limits _{0}^{ \infty } d \chi_\text{L}
e^{ - \chi_\text{L} C  }
\int\limits _{- \chi_\text{R}}^{  \chi_\text{L} } d \xi_\text{R}
\nonumber \\
&&
\times
e^{ - \xi_\text{R} \frac{ 1 - C }{2 } }
e^{ -i \xi_\text{R}^{2}  \frac{\zeta_\text{R} - C^{2} \zeta_\text{L} }{ 4}  }
e^{- i \xi_\text{R} \frac{ C x_\text{L}  }{ 2   }   }
e^{i  \chi_\text{L} \xi_\text{R} \frac{  \zeta_\text{R} - C^{2} \zeta_\text{L}  }{ 2   }   }
.
\label{b09}
\end{eqnarray}
Further, we split the interval of integration over $ \xi_\text{R}$ into two intervals, one from $0$ to $ \chi_\text{L}$ and the other from $ - \chi_\text{R}$ to $0$, and  represent $I_{3} = I_{3,1} + I_{3,2}$.

\subsubsection{The term $I_{3,1}$}

The first contribution is the following
\begin{eqnarray}
I_{3,1} &=& C
\int\limits _{0}^{ \infty } d \chi_\text{R}
e^{ - \chi_\text{R}  }
\int\limits _{0}^{ \infty } d \chi_\text{L}
e^{ - \chi_\text{L} C  }
\int\limits _{0}^{  \chi_\text{L} } d \xi_\text{R}
\nonumber \\
&&
\times
e^{ - \xi_\text{R} \frac{ 1 - C }{2 } }
e^{ -i \xi_\text{R}^{2}  \frac{\zeta_\text{R} - C^{2} \zeta_\text{L} }{ 4}  }
e^{- i \xi_\text{R} \frac{ C x_\text{L}  }{ 2   }   }
e^{i  \chi_\text{L} \xi_\text{R} \frac{  \zeta_\text{R} - C^{2} \zeta_\text{L}  }{ 2   }   }
.
\label{b10}
\end{eqnarray}
We integrate over $ \chi_\text{R}$, change the order of integration,
\begin{eqnarray}
\int\limits _{0}^{ \infty } d \chi_\text{L}
\int\limits _{0}^{  \chi_\text{L} } d \xi_\text{R}
\Rightarrow
\int\limits _{0}^{  \infty } d \xi_\text{R}
\int\limits _{ \xi_\text{R} }^{ \infty } d \chi_\text{L} ,
\label{b11}
\end{eqnarray}
and integrate over $ \chi_\text{L}$.
After these steps, we arrive at the following,
\begin{eqnarray}
I_{3,1} =
\int\limits _{0}^{  \infty } d \xi_\text{R}
\frac{
e^{ - \xi_\text{R} \frac{ 1 + C }{2 } }
e^{ i \xi_\text{R}^{2}  \frac{\zeta_\text{R} - C^{2} \zeta_\text{L} }{ 4}  }
e^{- i \xi_\text{R} \frac{ C x_\text{L}  }{ 2   }   }
}{ 1 - i   \xi_\text{R} \frac{  \zeta_\text{R} - C^{2} \zeta_\text{L}  }{ 2  C }  }
.
\label{b12} \\
\nonumber
\end{eqnarray}

\subsubsection{The term $I_{3,2}$}

In the second term, we change the sign of $ \xi_\text{R}$ and after  transformations similar to the ones above, we find
\begin{eqnarray}
I_{3,2} = C
\int\limits _{ 0 }^{ \infty } d \xi_\text{R}
\frac{
e^{  - \xi_\text{R} \frac{ 1 + C }{2 } }
e^{ -i \xi_\text{R}^{2}  \frac{\zeta_\text{R} - C^{2} \zeta_\text{L} }{ 4}  }
e^{i \xi_\text{R} \frac{ C x_\text{L}  }{ 2   }   }
}{
C + i \xi_\text{R} \frac{  \zeta_\text{R} - C^{2} \zeta_\text{L}  }{ 2   }
} .
\label{b13}
\end{eqnarray}
It is obvious that  $I_{3,2} = \left(  I_{3,1} \right)^{*}$, therefore, $I_{3} = 2 {\rm Re} I_{3,1}$.
Comparing Eq.~(\ref{b06}) and the double real part of Eq.~(\ref{b12}), we arrive at $I_{2} = - I_{3}$, as expected.

\section{The overlap integral $J_{+}$ for $ \psi_{1}$ from  Eq.~(\ref{26}) and $ \psi_{2}$ from Eq.~(\ref{27}) of the main text}
\label{appC}

Notice some change of integration variables compared to original Eqs.~(\ref{26}) and (\ref{27}) of the main text.
In particular, in $ \psi_{1}$ we use $ \chi_\text{R}$ instead of $ \xi$ and $ \psi_{2}$ is used in the form given in Eq.~(\ref{b00}).

We set $ \eta_\text{L} = \eta_\text{R} = +1$ and $ \tau_\text{R} = 0$, and represent the overlap integral defined in Eq.~(\ref{06}) of the main text as the sum of two terms, $J_{+} = J_{1+} +  J_{2+}$, where $\left(  x = t/ \Gamma _\text{L}, x_\text{L} = \tau_\text{L}/ \Gamma _\text{L} \right)$,
\begin{eqnarray}
J_{1+} &=&
\frac{ \sqrt{C} }{ \pi  }
\int\limits _{ - \infty}^{ \infty } dx
\int\limits_{0}^{ \infty} d \chi_\text{R}
e^{ -   \chi_\text{R}  }
e^{   i \chi_\text{R} x C    }
e^{ -i \chi_\text{R}^{2} \zeta_\text{R}  }
\int\limits_{0}^{ \infty} d \xi_{}
e^{ -   \xi_{}  }
e^{  - i \xi_{}   x        }
e^{  i \xi_\text{L}  x_{}       }
e^{ i \xi_{}^{2} \zeta_\text{L}  }
\nonumber \\
&=&
2 \sqrt{C}
\int\limits_{0}^{ \infty} d \chi_\text{R}
e^{ -   \chi_\text{R} \left(  1 + C \right)  }
e^{ -i \chi_\text{R}^{2} \left(  \zeta_\text{R} - C^{2} \zeta_\text{L} \right) }
e^{  i \chi_\text{R} C  x_\text{L}       } ,
\label{c01}
\end{eqnarray}
and
\begin{eqnarray}
J_{2+} &=&
- \frac{ \sqrt{C} }{2 \pi  }
\int\limits _{ - \infty}^{ \infty } dx
\int\limits_{0}^{ \infty} d \chi_\text{R}
e^{ -   \chi_\text{R}  }
e^{   i \chi_\text{R} x C    }
e^{ -i \chi_\text{R}^{2} \zeta_\text{R}  }
\nonumber \\
&&
\times
\int\limits _{0}^{ \infty } d \xi_\text{L}
e^{ - \frac{ \xi_\text{L} }{2 } }
e^{-i \xi_\text{L} x \frac{  1   }{ 2   }   }
e^{ i \xi_\text{L} \frac{  x_\text{L}  }{ 2   }   }
e^{ i \xi_\text{L}^{2}  \frac{\zeta_\text{L}  }{ 4}  }
\nonumber \\
&&
\times
\int\limits _{0}^{ \infty } d \xi_\text{R}
e^{ - \frac{ \xi_\text{R} }{2 } }
e^{-i \xi_\text{R} x \frac{ C }{ 2   }   }
e^{ i \xi_\text{R}^{2}  \frac{\zeta_\text{R}  }{ 4}  }
e^{i  \xi_\text{L} \xi_\text{R}  \frac{  \zeta_\text{R}  }{ 2 C   }   } .
\label{c02}
\end{eqnarray}
To evaluate $J_{2+}$, first we integrate over $x$.
This results in $2 \pi \delta \left(  C \chi_\text{R} - \xi_\text{L} - \xi_\text{R} C \right)$.
Then, we integrate over $ \xi_\text{L}  = C \chi_\text{R} - C \xi_\text{R} > 0$ preserving  $ \chi_\text{R} > \xi_\text{R}$.
Further, we introduce new variables, $ \bar x = \left(  \chi_\text{R} + \xi_\text{R} \right)/2$ and $ x = \chi_\text{R} - \xi_\text{R}$, transforming the area of integration as
\begin{eqnarray}
\int _{ \xi_\text{R} }^{ \infty } d \chi_\text{R}
\int _{0}^{ \infty } d \xi_\text{R} \Rightarrow
\int _{ 0}^{ \infty } d x
\int _{ x/2}^{ \infty } d \bar x
,
\label{c03}
\end{eqnarray}
integrating over $\bar x$, we get
\begin{eqnarray}
J_{2+} =
-2\sqrt{C}
\int\limits_{0}^{ \infty} d x
e^{ -x \left( 1 +  C \right)  }
e^{- i x^{2} \left(  \zeta_\text{R} - \zeta_\text{L} C^{2} \right)   }
e^{  i x C x_\text{L}   }
.
\label{c04}
\end{eqnarray}
Comparing the equation above and  Eq.~(\ref{c01}) we see that $J_{1+} + J_{2+} = 0$.
Therefore, indeed the two wave functions are orthogonal to each other.
In the same way one can show that $ \psi_{1}$, Eq.~(\ref{26}) of the main text, and $ \psi_{2}$, Eq.~(\ref{27}) of the main text, are orthogonal in the regime of a hole injection, $ \eta_\text{L} = \eta_\text{R} = -1$.

\section{The overlap integral $J_{-}$ for $ \psi_{1}$ from  Eq.~(\ref{26}) and $ \psi_{2}$ from Eq.~(\ref{27}) of the main text}
\label{appD}

Here we set $ \eta_\text{L} = -1$, $\eta_\text{R} = +1$, and $ \tau_\text{R} = 0$.
In $ \psi_{1}$ we use $ \chi_\text{R}$ instead of $ \xi$.
Now the function $ \psi_{2}$  looks as follows,
\begin{eqnarray}
\psi_{2}\left(  t \right) &=&
\frac{  1 }{  \sqrt{\pi  \Gamma _\text{L} } }
\int\limits_{0}^{ \infty} d \xi
e^{ -   \xi  }
e^{  i \xi \frac{ t - t_\text{L} }{ \Gamma _\text{L}  }    }
e^{ -i \xi^{2} \zeta_\text{L}  }
-\frac{  1 }{  2 \sqrt{\pi  \Gamma _\text{L} } }
\int\limits _{0}^{ \infty } d \xi_\text{L}
e^{ - \frac{ \xi_\text{L} }{2 } }
e^{i \xi_\text{L} \frac{  t_{} - t_\text{L}  }{ 2 \Gamma _\text{L}  }   }
e^{ - i \xi_\text{L}^{2}  \frac{\zeta_\text{L}  }{ 4}  }
\nonumber \\
&&
\times
\int\limits _{0}^{ \infty } d \xi_\text{R}
e^{ - \frac{ \xi_\text{R} }{2 } }
e^{-i \xi_\text{R} \frac{  t_{} }{ 2 \Gamma _\text{R}  }   }
e^{ i \xi_\text{R}^{2}  \frac{\zeta_\text{R}  }{ 4}  }
e^{-i  \xi_\text{L} \xi_\text{R}  \frac{  \zeta_\text{R}  }{ 2 C   }   }
.
\label{d00}
\end{eqnarray}
The overlap integral defined in Eq.~(\ref{06}) of the main text is represented as the sum of two terms, $J_{-} = J_{1-} +  J_{2-}$, where ($x = t/ \Gamma _\text{L}$,  $x_\text{L} = \tau_\text{L}/ \Gamma _\text{L}$, $C = \Gamma _\text{L}/ \Gamma _\text{R}$),
\begin{eqnarray}
J_{1-} &=&
\frac{ \sqrt{C} }{ \pi  }
\int\limits _{- \infty}^{ \infty } dx
\int\limits_{0}^{ \infty} d \chi_\text{R}
e^{ -   \chi_\text{R}  }
e^{   i \chi_\text{R} x C   }
e^{ -i \chi_\text{R}^{2} \zeta_\text{R}  }
\nonumber \\
&&
\times
\int\limits_{0}^{ \infty} d \xi_{}
e^{ -   \xi_{}  }
e^{  i \xi_\text{L} \left(  x - x_\text{L} \right)  }
e^{ -i \xi_{}^{2} \zeta_\text{L}  } = 0,
\label{d01}
\end{eqnarray}
and
\begin{eqnarray}
J_{2-} &=&
\frac{ 2 \sqrt{C} }{ \pi  }
\int\limits _{- \infty}^{ \infty } dx
\int\limits_{0}^{ \infty} d \chi_\text{R}
e^{ -   \chi_\text{R}  }
e^{   i \chi_\text{R} x C   }
e^{ -i \chi_\text{R}^{2} \zeta_\text{R}  }
\nonumber \\
&&
\times
\int\limits _{0}^{ \infty } d \xi_\text{L}
e^{ - \xi_\text{L}  }
e^{i \xi_\text{L} \left(  x_{} - x_\text{L} \right)    }
e^{ -i \xi_\text{L}^{2} \zeta_\text{L}   }
\int\limits _{0}^{ \infty } d \xi_\text{R}
e^{ - \xi_\text{R} }
e^{-i \xi_\text{R}  x_{} C   }
e^{ i \xi_\text{R}^{2}  \zeta_\text{R}    }
e^{-i  2 \xi_\text{L} \xi_\text{R}  \frac{  \zeta_\text{R}  }{  C   }   } .
\label{d02}
\end{eqnarray}
The integration over $x$ results in $2 \pi \delta\left( \chi C +    \xi_\text{L}  -  \xi_\text{R} C\right) $.
Integrating subsequently over $ \xi_\text{L}$, we take into account that $ \xi_\text{L}$ is positive: $\xi_\text{L} =   \xi_\text{R} C -  \chi C >0 \Rightarrow \xi_\text{R} > \chi $.
Then we make a shift $ \xi_\text{R} \to \xi_\text{R} + \chi$, and integrate over $ \chi$,
\begin{eqnarray}
J_{2-} =
2 \sqrt{C}
\int\limits _{ 0}^{ \infty } d \xi_\text{R}
e^{ -  \xi_\text{R} \left(  1+C \right) }
e^{ -i  \xi_\text{R}^{2}  \left(  \zeta_\text{R} + \zeta_\text{L} C^{2} \right)    }
e^{-i  \xi_\text{R}  C  x_\text{L}    }
.
\label{d03}
\end{eqnarray}
The equation above is nothing but Eq.~(\ref{28}) of the main text with $ \xi = \xi_\text{R}\left(  1+C \right)$ and $ \eta_{j}$'s introduced.


%

%


\end{document}